# Jefferson Lab
## Thomas Jefferson National Accelerator Facility

Physics Opportunities with
the 12 GeV Upgrade at Jefferson Lab

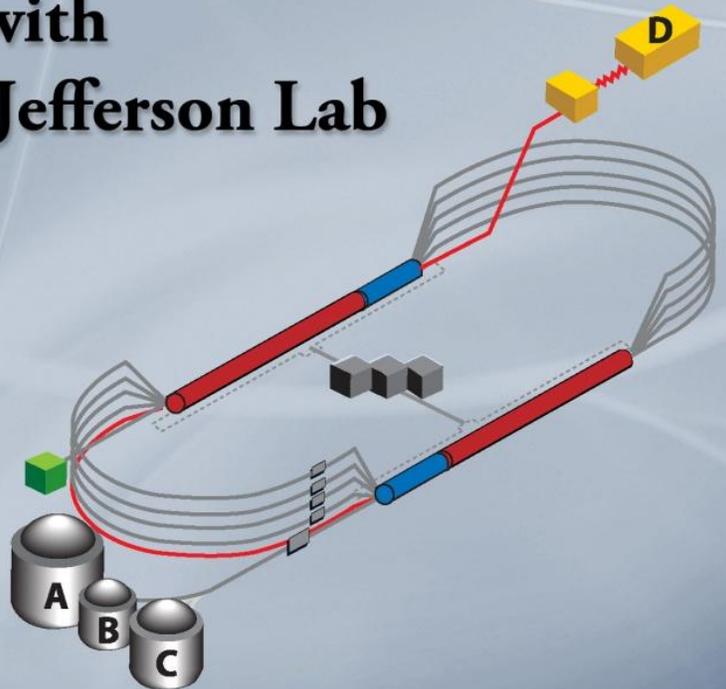

**Jozef Dudek**
*Thomas Jefferson National Accelerator Facility, Newport News, VA 23606 USA*
*Old Dominion University, Norfolk, VA 23529*

**Rolf Ent**
*Thomas Jefferson National Accelerator Facility, Newport News, VA 23606 USA*

**Rouven Essig**
*C.N. Yang Institute for Theoretical Physics, Stony Brook, NY 11794*

**Krishna Kumar**
*University of Massachusetts, Amherst, MA 01003*

**Curtis Meyer**
*Carnegie Mellon University Pittsburgh, PA 15213*

**Robert McKeown**
*Thomas Jefferson National Accelerator Facility, Newport News, VA 23606 USA*

**Zein Eddine Meziani**
*Temple University, Philadelphia, PA 19122*

**Gerald A. Miller**
*University of Washington, Seattle, WA 98195*

**Michael Pennington**
*Thomas Jefferson National Accelerator Facility, Newport News, VA 23606 USA*

**David Richards**
*Thomas Jefferson National Accelerator Facility, Newport News, VA 23606 USA*

**Larry Weinstein**
*Old Dominion University, Norfolk, VA 23529*

**Glenn Young**
*Thomas Jefferson National Accelerator Facility, Newport News, VA 23606 USA*

**Susan Brown, Administrative Support**
*Thomas Jefferson National Accelerator Facility, Newport News, VA 23606 USA*

# Table of Contents

# 1 Executive Summary

## Overview

We are at the dawn of a new era in the study of hadronic nuclear physics. The non-Abelian nature of Quantum Chromodynamics (QCD) and the resulting strong coupling at low energies represent a significant challenge to nuclear and particle physicists. The last decade has seen the development of new theoretical and experimental tools to quantitatively study the nature of confinement and the structure of hadrons comprised of light quarks and gluons. Together these will allow both the spectrum and the structure of hadrons to be elucidated in unprecedented detail. Exotic mesons that result from excitation of the gluon field will be explored. Multidimensional images of hadrons with great promise to reveal the dynamics of the key underlying degrees of freedom will be produced. In particular, these multidimensional distributions open a new window on the elusive spin content of the nucleon through observables that are directly related to the orbital angular momenta of quarks and gluons. Moreover, computational techniques in Lattice QCD now promise to provide insightful and quantitative predictions that can be meaningfully confronted with, and elucidated by, forthcoming experimental data. In addition, the development of extremely high intensity, highly polarized and extraordinarily stable beams of electrons provides innovative opportunities for probing (and extending) the Standard Model, both through parity violation studies and searches for new particles. Thus the 12 GeV upgrade of the Continuous Electron Beam Accelerator Facility (CEBAF) at Jefferson Lab will enable a new experimental program with substantial discovery potential to address these and other important topics in nuclear, hadronic and electroweak physics.

## A New Experimental Vista

The upgrade of CEBAF and associated experimental equipment at Jefferson Lab is presently underway, with completion expected in FY15. The upgraded facility will accelerate electron beams to 11 GeV for experiments in the existing Halls A, B and C. In addition, a12 GeV beam can be provided to a new experimental area to generate a 9 GeV tagged photon beam, enabling a powerful program of meson spectroscopy in Hall D. The facility is capable of delivering beam to any 3 of the 4 halls simultaneously. Appendix A describes the planned experimental equipment in each of the 4 halls.

The capabilities of the upgraded CEBAF will represent a significant leap over previous technology, with an unmatched combination of beam energy, quality and intensity. Although lower luminosity beams of electrons and photons in the energy range of the upgraded CEBAF have been previously available, the new facility at Jefferson Lab capitalizes on remarkable developments in experimental techniques over the last decade. Even at the advent of CEBAF operations in 1995, one could only dream of the availability of highly polarized (86%) continuous wave (CW) beam with intensities up to 180 $\mu$A, as have been delivered during the last year. The revolutionary construction and operation of the CEBAF Large Acceptance Spectrometer (CLAS) has proved that large acceptance devices can be successfully operated in the environment of an external CW electron beam, and the advent of new detector technologies and high rate electronics offer a powerful combination of experimental tools that far exceed the capability of previous experiments and facilities.

During the last few years, the Jefferson Lab Program Advisory Committee (PAC) has considered experimental proposals from the Jefferson Lab user community (over 1300 scientists) for the upgraded facility. The PAC was charged to only approve proposals that were considered to be rated in the top 50% of anticipated experiments, so the approved program represents an above average standard. Interest in the 12 GeV science program has been tremendous, and there are presently 52 approved proposals, each with a scientific rating and recommended allocation of beam time (see Appendix B for a list). This set of approved experiments will require more than 6 years of running the upgraded facility at near-full efficiency, extending scientific productivity well into the 2020's. The PAC continues to meet on an annual basis to consider new proposals. In addition, the user community is considering various additions to the experimental equipment under construction, and these are also briefly discussed in Appendix A.

This White Paper provides an up-to-date description of the science program and potential of the upgraded facilities at Jefferson Lab. It is primarily based on material in the experiment proposals already approved by the PAC. In this Overview, we present a brief introduction to the overall scope and major themes of the anticipated program. The following sections discuss the four main science thrusts in more detail.

## Science Opportunities

The implementation of a new meson spectroscopy program in the mass range up to 3.5 GeV will offer insight into the role of gluon self interactions and the nature of confinement. Previous models of undiscovered mesons with exotic quantum numbers are now complemented by robust Lattice QCD calculations that indeed predict these states within the mass range accessible to the *GlueX* experiment being constructed in Hall D. *The prospects for discovering these mesons have increased dramatically in the past few years.* In addition, the detailed spectroscopic information from experiment, coupled with the guidance of new Lattice QCD results, offers an exciting and unique opportunity to explore mechanisms of confinement.

The study of the internal landscape of the nucleons is now undergoing a renaissance. Driven by the inadequacies of previous treatments and by recent experimental data, we are now moving beyond the simple one dimensional parton distribution functions of the past. The pioneering efforts of HERMES and COMPASS, together with the 6 GeV Jefferson Lab, have demonstrated the feasibility of studying Transverse Momentum Distributions (TMDs) as well as Deeply Virtual Compton Scattering (DVCS) measurements that offer access to Generalized Parton Distributions (GPDs). Indeed, recent measurements using the 6 GeV CEBAF have demonstrated that high quality CW polarized electron beams with a combination of large acceptance and precision detectors are powerful tools to attack these new observables. We can now be confident that the extended kinematic range and new experimental hardware associated with the Jefferson Lab 12 GeV Upgrade will provide access to these fundamental underlying distributions and reveal new aspects of nucleon structure. *It is quite possible that much of the remaining nucleon spin will be found in the orbital motion of the valence quarks.*

QCD suggests the existence of novel phenomena in nuclear physics. The nuclear medium provides mechanisms for filtering quantum states and studying their spacetime evolution. Quantum fluctuations in nuclei cause local high-density regions of cold nuclear matter, up to four times larger than typical nuclear densities, comparable to neutron star core densities. Recent Jefferson Lab experiments have shown that these fluctuations are related to the nuclear quark distributions. *Future experimental studies of the fundamental short-distance properties in nuclei will provide a quantitative understanding of nuclear properties and their relation to the parton distributions in nuclei.* Important studies of nuclear structure physics will also be possible through, for example, the use of parity violation to precisely determine the neutron skin of heavy nuclei.

Over the last few years we have seen a remarkable growth in the proposed program to test the Standard Model of electroweak interactions using the upgraded CEBAF. While the extension of the parity violation program was indeed envisioned as part of the upgrade plan, we can now foresee a powerful program to provide higher precision tests of the Standard Model and learn about the unseen forces that were present at the dawn of the universe. These experiments push the technology at Jefferson Lab to the limits, and recent success with the Qweak experiment clearly points the way to a more ambitious program in the future. In addition, we have received new proposals to search for heavy neutral vector bosons that could explain the muon *g-2* anomaly as well as offer connections to theories of Dark Matter. *The upgraded facilities at Jefferson Lab will provide exceptional opportunities for discovery of new phenomena beyond the Standard Model.*

In the next sections we will review the future Jefferson Lab experimental program that, when combined with parallel theoretical studies, will elucidate how QCD and the confinement of quarks and gluons determine the detailed structure and dynamics of baryons, mesons, and all nuclear matter, as well as extend the sensitivity to new physics in the electroweak sector.

## International Context

Jefferson Lab is a remarkable and unique facility, but there are also complementary efforts on the international scene that provide a broader context for the Jefferson Lab program. COMPASS at CERN continues with large acceptance at much lower luminosity, but higher center-of-mass energies. COMPASS can generally provide similar information to Jefferson Lab, but with lower statistical precision and at lower Bjorken *x* ($x_B$). (COMPASS has just received approval for several years of further running at CERN.) Mainz has a 2 GeV CW microtron facility, with excellent CW polarized electron beams. (This facility has just received renewed multiyear funding from the German government.) However, the lower energy at Mainz implies a very limited kinematic reach relative to Jefferson Lab.

Hadron beam facilities that provide relevant experimental capabilities are also coming on line. JPARC in Japan will explore strange hadronic systems with high intensity kaon and pion beams. PANDA at GSI in Germany will study meson spectroscopy in proton-antiproton collisions focusing on the charmed-quark sector that will be very complementary to the *GlueX* program at Jefferson Lab. Naturally we are seeing a strong interest in mutual collaboration on meson spectroscopy from GSI and other German institutes. In addition, there are strong connections between the Jefferson Lab program and the Minerva and Drell-Yan studies at Fermilab.

Jefferson Lab should be seen as an integral component of a worldwide program. There is indeed a large and active international community studying QCD, confinement, and precision tests of the Standard Model. Jefferson Lab is clearly providing the US nuclear physics program with a leadership role in this vibrant international community, and with the successful upgrade, will be a flagship for this diverse scientific program for many years to come.

# 2 Meson Spectroscopy, Hybrid Mesons & Confinement

In this section, we will review the future Jefferson Lab experimental program that aims to understand the nature of gluonic fields in the structure of hadrons, and to search for the effects of excitations of the gluonic field on the observed spectrum of mesons. When combined with parallel theoretical studies, these experiments will elucidate how QCD, with its confinement of quarks and gluons, determines the detailed structure and dynamics of hadrons. The 12 GeV Upgrade will, with its new experimental Hall D receiving a dedicated photon beam, provide a unique capability to study the inner workings of meson states and conduct a definitive search for excited hadrons whose quantum numbers are determined by gluonic degrees of freedom. In addition, new tests of predictions of chiral perturbation theory will become possible through the study of decays of the η meson.

## 2a QCD and the Meson Spectrum

The hadron spectrum offers a window on the properties of QCD in its domain of strong-coupling, where non-perturbative effects can give rise to dramatic phenomena that cannot be naively read off from the quark-gluon Lagrangian. The most striking manifestation of non-perturbative dynamics in QCD is also the least well understood: the confinement of quarks and gluons within hadrons. Unlike in atomic physics where electron excitation will eventually lead to ionization, the excited state spectrum of QCD continues to be expressed solely in terms of hadrons and never free quarks or gluons. Investigation of the systematics of the excited spectrum of hadrons may prove to be the ideal tool to determine the nature of the confining mechanism of QCD.

A description of the gross features of the observed meson spectrum begins with noting that the relative lightness of the pion suggests a spontaneous breaking of the approximate chiral symmetry present for light *up* and *down* quarks. The same non-perturbative physics is believed to cause the dressing of these very light quarks into quasi-particle 'constituent' quarks with mass scale in the hundreds of MeV region. Such effective degrees-of-freedom were proposed long

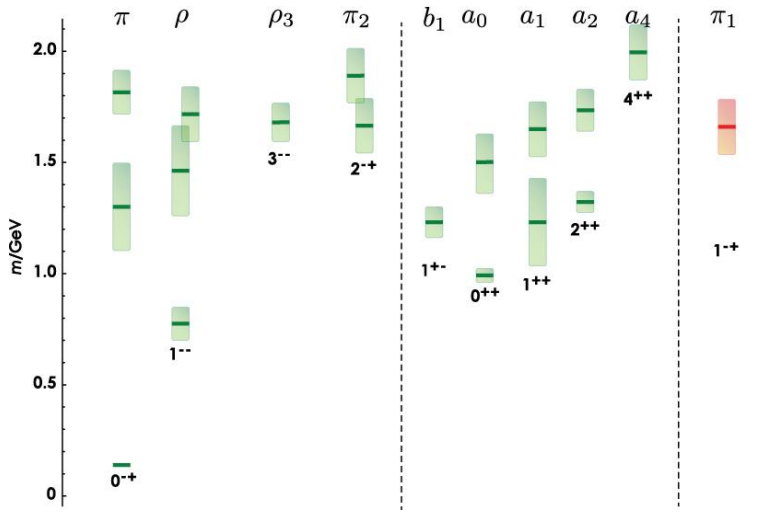

*Figure 2.1: The experimental spectrum of isovector mesons as summarized by the Particle Data Group [2-1]. The vertical height of each box indicates the hadronic decay width of each state. The rightmost column shows the controversial exotic candidate state, $\pi_1(1600)$.*

before QCD as a way to describe the systematics of the observed meson (and baryon) spectra. The experimental spin and parity ($J^{PC}$) distribution of states, and the lack of meson states with isospin or strangeness beyond one unit, suggests that mesons can be described simply in terms of a constituent quark and antiquark. The coupling of the spin-½ quark and antiquark, allowing for orbital angular momentum between them, yields a clear prediction of the spin and parity quantum numbers of the meson states. In nearly all the allowed cases, $J^{PC} = 0^{-+}, 0^{++}, 1^{--}, 1^{+-}, 1^{++}, 2^{--}, 2^{-+}, 2^{++}, ...,$ candidate states have been observed experimentally *(Figure 2.1)*

We see that the role of the gluonic field in determining the spectrum of hadrons has been partially exposed - it dresses the bare quarks to give the constituent quarks whose excitations of motion determine the low-lying meson spectrum. But this surely is not the only role the gluonic field plays within the hadron spectrum. As a strongly coupled system the gluonic field should have its own excitation spectrum that should manifest itself as states beyond the $q\bar{q}$ constructions discussed above. Even in a theory without quarks, there would exist bound-states made purely of glue, the 'glueballs'. With quarks present, the glueball basis states can mix strongly with isoscalar $q\bar{q}$ states of the same $J^{PC}$ and this has stymied attempts to extract clear information about glueballs from the experimental meson spectrum [2-2, 2-3].

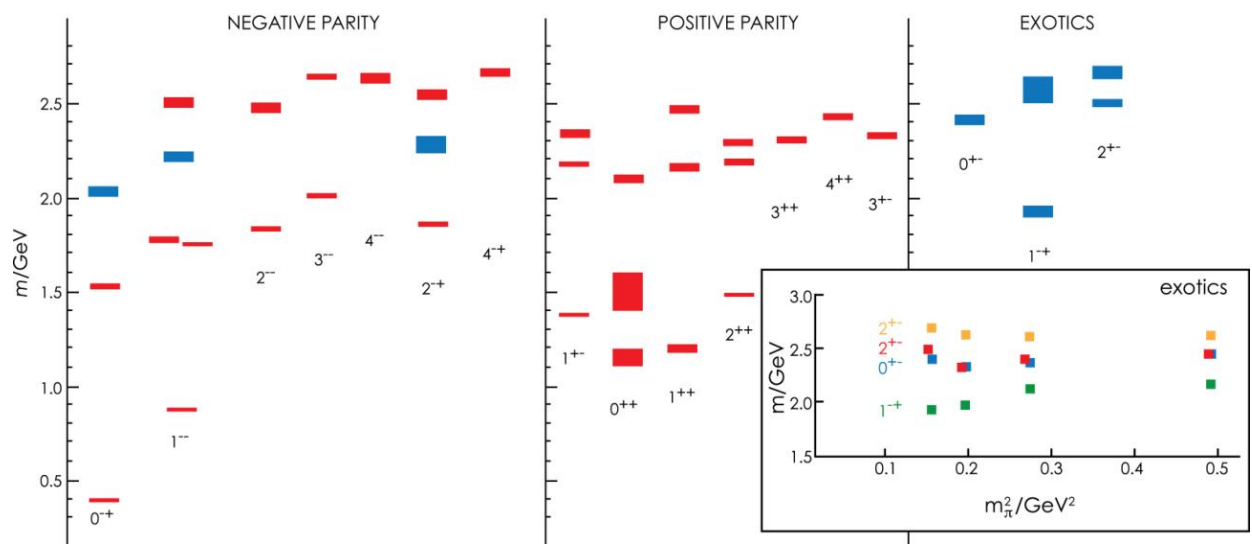

*Figure 2.2: The spectrum of isovector mesons extracted from a lattice QCD calculation [2-4, 2-5, 2-6, 2-7] with light quark masses corresponding to $m_\pi \sim 400$ MeV. A clear spectrum of $q\bar{q}$ excitations (red) is supplemented with a spectrum of exotic and non-exotic hybrid mesons (blue). The vertical height of each box indicates the statistical uncertainty on the mass within the calculations – the hadronic widths are not determined. Inset: the quark mass dependence of the exotic meson spectrum demonstrating the robustness of the calculated observables.*

A more promising sector has an excited gluonic field coupled to a $q\bar{q}$ pair, with states of this type being known as *hybrid mesons*. If the gluonic field excitation has quantum numbers other than $0^{++}$ we have the possibility that when coupled to the quantum numbers of the $q\bar{q}$ pair we might generate meson $J^{PC}$ outside the set accessible to $q\bar{q}$ alone, e.g. $J^{PC} = 0^{+-}, 0^{--}, 1^{-+}, 2^{+-}$. These are known as *exotic $J^{PC}$* and observing a state with these quantum numbers would be a smoking gun signature for physics beyond the $q\bar{q}$ picture and would suggest a non-trivial role for gluonic excitations in the hadron spectrum.

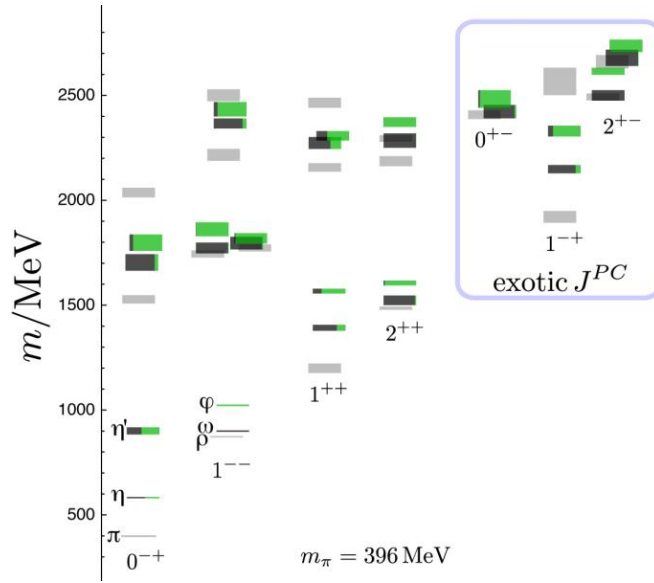

*Figure 2.3 Partial spectrum of isoscalar mesons extracted from a lattice QCD calculation [2-7] with light quark masses corresponding to $m_\pi \sim 400$ MeV. Green/Black boxes show isoscalar states with green indicating $s\bar{s}$ and black indicating $u\bar{u}+d\bar{d}$ content. Grey boxes show the isovector spectrum for comparison. The experimentally determined mixing pattern is reproduced for conventional states with $J^{PC} = 0^{-+}, 1^{--}, 1^{++}, 2^{++}$ and the mixing pattern for exotic states, to be measured in GlueX, is predicted.*

As well as *exotic* hybrids, we also anticipate there being hybrid mesons with *non-exotic* spin-parity quantum numbers. A more detailed understanding of the nature of the gluonic excitation will follow if we can identify these states, embedded within the spectrum of conventional $q\bar{q}$ states. Recent theoretical work [2-4, 2-5, 2-6, 2-7] using Lattice QCD methods has identified the lowest-lying set of isovector hybrid mesons (**Figure 2.2**), with non-exotic $J^{PC}= 0^{-+}, 1^{--}, 2^{-+}$ states partnering a lightest exotic $1^{-+}$ state at an energy scale roughly 1.3 GeV above the $\rho$ meson. An explanation of this spectrum is provided by the lowest-energy gluonic excitation being of chromomagnetic character - heavier exotic states, $0^{+-}, 2^{+-}$ are also observed, and appear to correspond to quark orbital excitation in addition to the same chromomagnetic gluonic excitation. (There are also strong suggestions that the same gluonic excitation plays the dominant role in determining the hybrid *baryon* spectrum [2-8]). The spectrum shown in (**Figure 2.2**) is obtained from a calculation where the light-quark masses are larger than their true physical values, but the qualitative features of the hybrid spectrum do not appear to change as the light quark mass is varied, as shown in the inset in **Figure 2.2**. Computations at lighter quark masses will be required to confidently estimate the mass scale, but it is clear from the trends observed that much, if not all, of this spectrum will be within the energy reach of the Jefferson Lab 12 GeV Upgrade in the *GlueX* experiment.

Recent lattice calculations [2-7] of the *isoscalar* mesons have provided estimates of the $(u, d, s)$ quark flavor composition of states (**Figure 2.3**). This addresses the question of why certain states like the ω and φ are ideally mixed by quark flavor ($u\bar{u}+d\bar{d}$ and $s\bar{s}$), while others like η and η′ correspond to a completely different admixture. As a result of the lattice calculations, we now have suggestions for the mixing patterns of hybrid mesons and these predictions can be experimentally tested at Jefferson Lab through study of the decay modes of the hybrid mesons. Empirically, the decay of $s\bar{s}$ states to kaon-pair final states are enhanced relative to non-strange decays. In order to be able to map this out for the hybrid states, it is necessary to be able to measure final states involving several particles, both charged and neutral, and identify kaons.

## 2b      The Experimental Program

High statistics data on hadron resonances observed in multiple final states using novel production methods can revolutionize spectroscopy as demonstrated by the recent charmonium 'renaissance'. Spectroscopy in the charmonium sector, which once appeared to be simply understood in terms of predictable $c\bar{c}$ bound states, has undergone an extraordinary transformation in the last ten years with the discovery of over a dozen new states by the *B*-factory and Tevatron experiments. While none of the observed states exhibit exotic $J^{PC}$ quantum numbers, several have rather unusual properties in comparison to the ordinary, established, $c\bar{c}$ mesons. This has spurred a flurry of theoretical activity in which previously assumed properties of non-perturbative QCD are being questioned.

The goal of *GlueX* is to provide a comparable enhancement in the status of the light meson spectrum by introducing the novel mechanism of *photoproduction*. There is little existing data on the photoproduction of light meson resonances, especially in the mass region where hybrid mesons are predicted to lie, and theoretical suggestions are that hybrid states may be preferentially produced with photons. Not only is this favored in a number of phenomenological calculations [2-9, 2-10, 2-11, 2-12], but lattice calculations that accurately reproduce the radiative decay of charmonium states also indicate that the radiative couplings of hybrid mesons are not small [2-13]. *GlueX* will use the coherent bremsstrahlung technique to produce a linearly-polarized real photon beam. A solenoid-based hermetic detector will collect data on meson resonance production through decays to charged and neutral final state particles. Statistics after the first full year of running will increase the current world photoproduction database from a few thousand events per final state to millions of events. This improvement by several orders of magnitude will provide a new window on mesons in the 1.8 to 2.7 GeV mass range.

Data supporting the existence of states with exotic $J^{PC}$ are still sparse. A recent review article [2-14] provides a summary of the field. Briefly, some tentative evidence exists for up to three isovector, $J^{PC}=1^{-+}$, exotic states: $\pi_1(1400)$, $\pi_1(1600)$ and $\pi_1(2015)$, although none of these are without controversy. These observations have been primarily made in hadron-beam experiments, often with relatively low statistics. *GlueX's* use of the novel photoproduction technique and collection of high statistics in a large number of hadronic final states, including those of high multiplicity, will remedy these shortcomings. Measurements with lower-multiplicity final states using nearly-real polarized photons will be made by CLAS12. These complementary studies will provide important cross-checks on observations made in *GlueX* as well as making some studies using virtual photons possible.

Photoproduction of a particular multi-meson final state will populate many partial waves simultaneously, and a spectrum of meson resonances only follows when the data is analyzed in terms of partial wave amplitudes. To be able to fully exploit these new data, and connect them to the theoretical predictions, extensive phenomenological work needs to be carried out to improve the current state of amplitude analysis. A key mission of the new Jefferson Lab Physics Analysis Center is to establish a state-of-the-art framework for such analyses, building on, and networking with, global theoretical and phenomenological expertise. The application of such techniques to forthcoming results from CLAS12 and *GlueX* on both meson and baryon photo and electroproduction is essential for laying the foundation for robust partial wave analyses, and so ensuring the capability of these experiments to establish the existence of states in which gluonic degrees of freedom are excited.

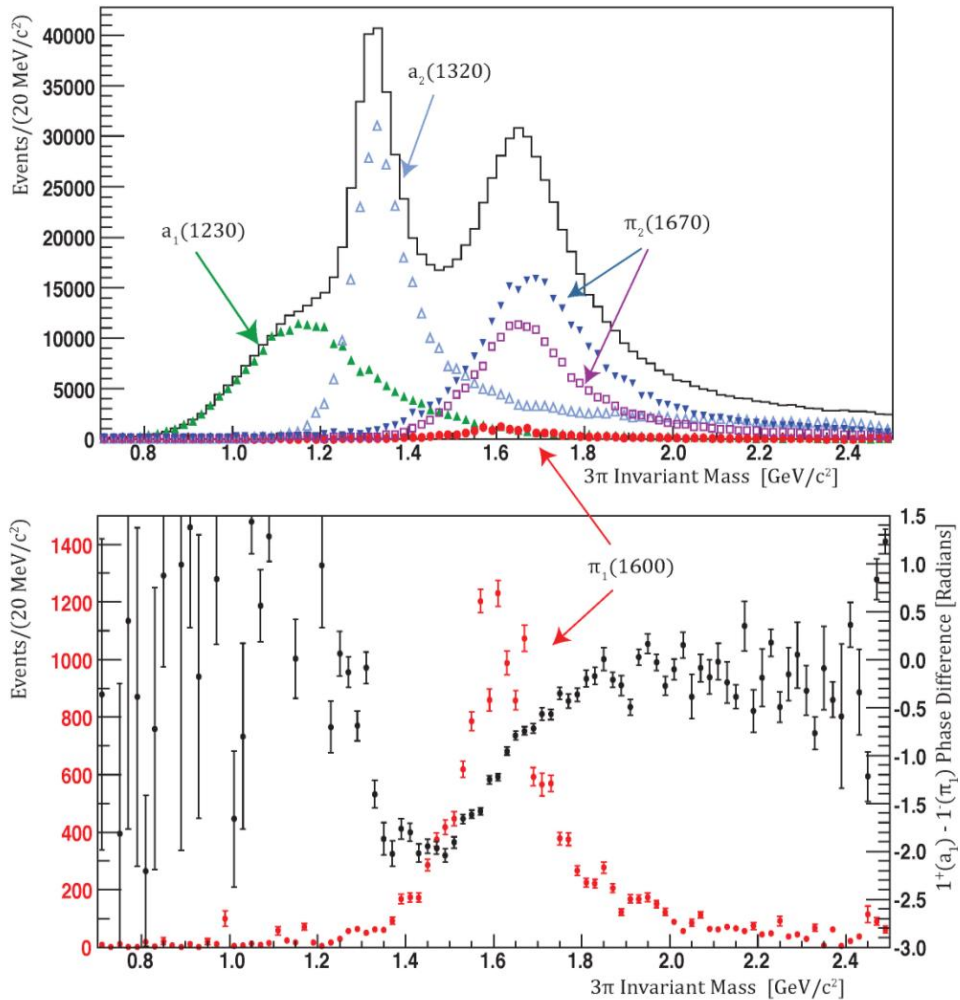

*Figure 2.4: An amplitude analysis carried out using the full GlueX software suite showing a small exotic signal being cleanly extracted from the much stronger conventional signals in the data. The top plot shows the total cross section (solid curve) and the extracted intensities of several partial waves. Of note is the reproduction of a very small signal for an exotic $\pi_1(1600)$. The strength of this wave, which may be large in actual photoproduction, is chosen to be small in the simulation to test the sensitivity of the analysis methodology. The bottom plot shows the weak exotic signal on a large scale as well as the observed phase motion between the signal and one of the stronger waves. Such an extracted phase motion would be clear evidence for resonant behavior of the signal.*

The ability of *GlueX* to extract even a small exotic signal from data which is dominated by conventional meson production and backgrounds has been explored in simulations of the response of the detector to a realistic distribution of final-state particles. Amplitude analysis of Monte Carlo simulated data which has passed through a detailed model of the *GlueX* detector is able to reproduce the input model of resonance production. As seen in **Figure 2.4** even an artificially small signal for exotic $\pi_1(1600)$ in $\gamma p \rightarrow \pi^+\pi^+\pi^- n$ is clearly extracted alongside the much larger production of conventional mesons. While the amplitude analysis model used to generate and analyze the simulated data does not yet incorporate all of the constraints required by QCD, this study demonstrates that the acceptance provided by the detector design does facilitate exotic searches even when the signals are significantly suppressed compared to dominant channels.

A thorough understanding of the spectrum of mesons in QCD demands complementary efforts in experiment, theory and phenomenology. *GlueX* is unique both in the size of its expected data set and in its optimization for production of exotic light-quark systems. Through the collection of high statistics in modern experiments, the previously simple view of charmonium has had to be overhauled to allow for the many new states observed. The 12 GeV physics program at Jefferson Lab offers the possibility to do the same for the light-quark sector, providing a window on the mechanisms by which the gluonic field of QCD confines light quarks in hadron resonances.

## 2c  Tests of Chiral Symmetry and Anomalies via the Primakoff Effect

The *GlueX* detector allows not only the detailed study of spectroscopy but, using the Primakoff effect, the measurement of the two photon decays of the $\eta$ and $\eta'$ mesons as well, with far greater precision than previously. Such measurements test our understanding of two basic phenomena of QCD, namely the spontaneous breaking of chiral symmetry and the role of chiral anomalies. Such anomalies are manifested in their most unambiguous form in the sector of light pseudoscalar mesons. The chiral anomaly induced by the electromagnetic interactions drive the two-photon decays of the $\pi^0, \eta,$ and $\eta'$, while the axial anomaly induced by QCD itself breaks the axial *U(1)* symmetry and generates a large fraction of the $\eta'$ mass.

The 12 GeV upgrade will make it possible to extend the current development of the recently completed high precision measurement of the $\pi^0 \rightarrow \gamma\gamma$ decay width via the Primakoff effect to include the $\eta$ and $\eta'$ mesons, and make possible a low $Q^2$ measurement of their transition form factors. Of particular importance is a more accurate determination of the quark-mass ratio that will result from a precise knowledge of the $\eta \rightarrow \pi^+\pi^-\pi^0$ decay rate, which in turn is determined from its well known branching fraction with respect to the rate for $\eta \rightarrow \gamma\gamma$. **Figure 2.5** illustrates the current situation where the ratio $\mathcal{R} = (m_s^2 - \hat{m}^2)/(m_d^2 - m_u^2)$ with $\hat{m}$ the mean *up* and *down* quark mass, is determined from kaon masses and the $\eta \rightarrow \pi^+\pi^-\pi^0$ rate, and the projected determination of $\mathcal{R}$ from the proposed precision Primakoff measurement of $\Gamma(\eta \rightarrow \gamma\gamma)$ [2-17]. This measurement will test our understanding of the essentials of chiral dynamics in the $\eta$ and $\eta'$ sector as never before.

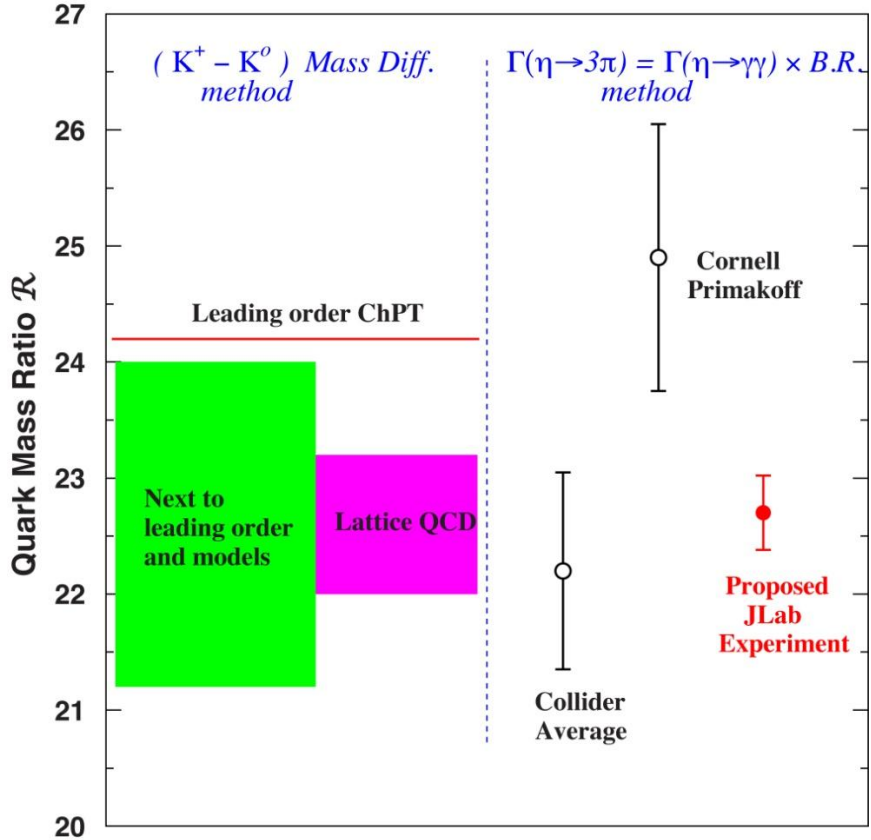

*Figure 2.5: Values of the quark mass ratio ℛ defined in the text: the red line denotes the result for ℛ using leading order Chiral Perturbation Theory with electromagnetic corrections to Kaon masses from Dashen's Theorem. The green box represents the corresponding result at next to leading order, which requires model inputs --- the size of the band delineates the uncertainties [2-18]. The purple box gives the result from Lattice QCD with the electromagnetic corrections to Kaon masses from Ref. [2-15]. These are to be compared with data on the right. These data result from the extractions of ℛ from experimental information on $\eta \rightarrow 3\pi$ from the Cornell Primakoff measurement and from electron-positron colliders as analyzed in Refs. [2-16]. The uncertainty expected from the proposed experiment at Jefferson Lab is E12-10-011 [2-17] is indicated by the red datapoint.*

# 3    The Internal Structure of Hadrons

Nucleons and their composite nuclei constitute almost all of the known visible matter in the universe, yet we lack a detailed understanding of these objects from first principles. Our grasp of the fundamental quark and gluon structure of the nucleon founded on QCD is still in its infancy compared to that of atomic structure based on quantum electrodynamics (QED). Nevertheless, in the last ten years we have entered a new era where a framework suitable for a comprehensive and quantitative approach to the description of nucleon structure has emerged [3-1]. In this framework our knowledge of nucleon structure is encoded in the Wigner distributions of the constituents, a quantum mechanical concept, introduced in 1932 [3-2], that is analogous to the classical concept of a phase space distribution. From the Wigner distributions, a natural interpretation of measured observables is provided through the construction of quantities known as generalized parton distributions (GPDs) and transverse momentum-dependent distributions (TMDs), shown in *Figure 3.1*. While GPDs are key to the "spatial" tomography of the nucleon, TMDs allow for its "momentum" tomography [3-3].

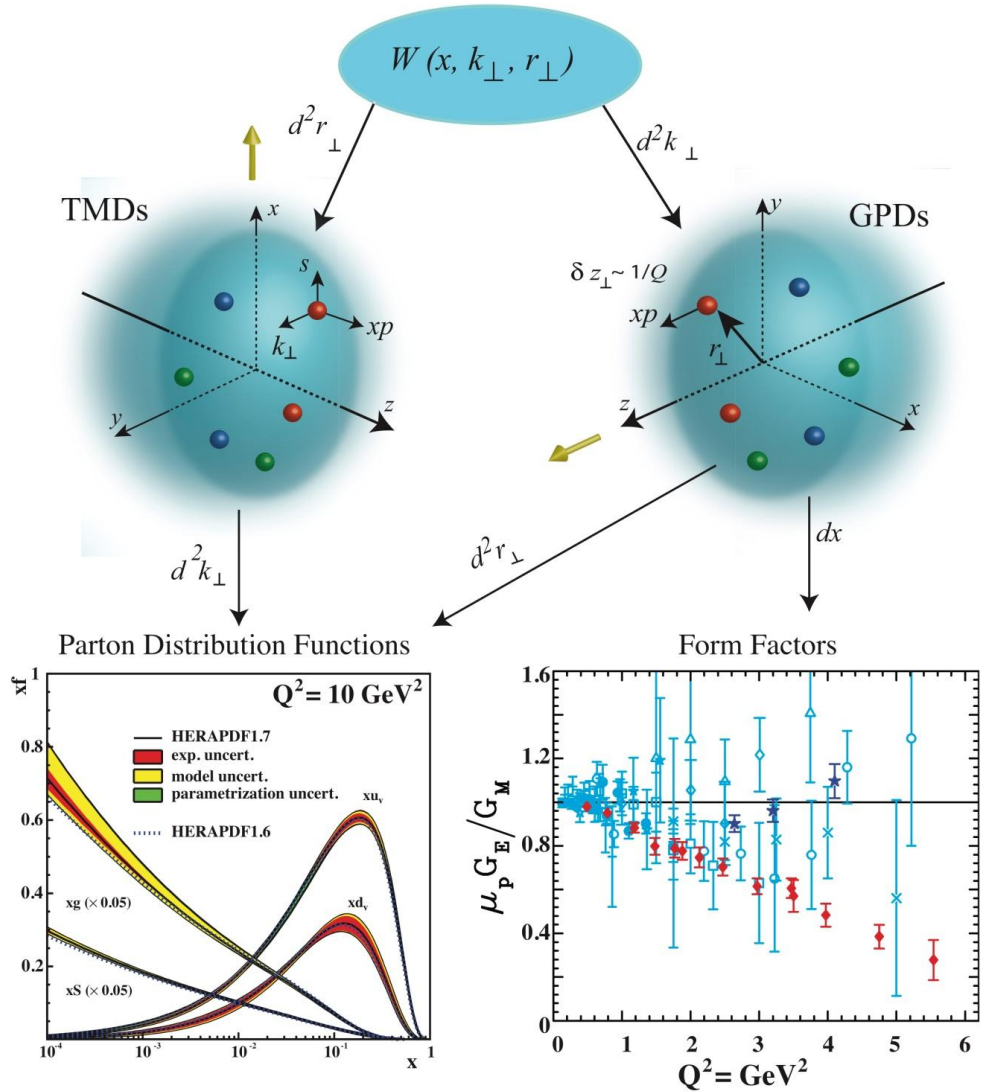

*Figure 3.1: The Wigner distributions yield a unified description of a nucleon in terms of the position and momenta of its constituents.  The uncertainty principle precludes knowing both position and momentum simultaneously, but the three-dimensional Generalized Parton Distributions (GPDs) and Transverse-Momentum-Dependent Distributions (TMDs) provide a powerful spatial and momentum tomography.  The differential variables along the arrows indicate the variable integrated over to move from the upper to lower distributions.*

Past studies of deep inelastic electron scattering offered us merely a one dimensional view of nucleon structure where we learned about the motion of partons parallel to the direction of travel of the nucleon. Their longitudinal momentum distribution is then described by parton distribution functions (PDFs). The nucleon was viewed as a collection of fast moving quarks, antiquarks and gluons, whose transverse momenta were not resolved. While simple and elegant, such a description is unable to address one of the key questions in our understanding of the nucleon, namely how its spin is apportioned between the spin of its constituents and their orbital angular momentum. To understand this requires a three-dimensional description.

A representation using GPDs and TMDs is driven by the overwhelming need to go beyond the one dimensional picture of nucleon structure [3-4]. Even at large Bjorken *x*, where most of the longitudinal momentum of the proton is carried by valence quarks, seemingly puzzling results from a first generation of worldwide exclusive and semi-inclusive deep inelastic lepton scattering experiments require a GPD and TMD description for their interpretation. Thus these representations provide us with a unified view, demonstrating both the importance of this new phenomenology, and the limitation of our previous studies of nucleon structure.

Knowledge from inclusive, semi-inclusive and exclusive electron scattering using the 12 GeV CEBAF upgrade will provide information on the transverse position and transverse momentum of quarks for a fixed slice of their longitudinal momentum leading to a three-dimensional imaging of the nucleon both in position and momentum. This upgrade offers for the first time the tools to unravel the nucleon valence quark structure by mapping the spatial position and momentum distribution of the quarks with sufficient precision to propel our knowledge and understanding of the basic building blocks of nuclear matter to a level unmatched previously.

## 3a  A multi-dimensional view of nucleon structure

The concept of Wigner distributions has inspired a description of the nucleon that goes beyond the one-dimensional pictures contained in the elastic form factors and the longitudinal parton distributions, yet encompasses them [3-2]. The Generalized Parton Distributions (GPDs) emerge after integrating the Wigner distributions over the transverse momentum and longitudinal position of partons, *Figure 3.1*. GPDs encode the correlation between the quark/gluon transverse position in the nucleon and its longitudinal momentum, and can be measured directly in exclusive scattering processes at large $Q^2$, in which the nucleon is observed intact in the final state, namely, deep virtual Compton scattering (DVCS) and deep virtual meson production (DVMP). They offer a path to a full 3-dimensional exploration of the nucleon structure, in transverse position and longitudinal momentum space, and thus provide spatial tomography of the nucleon.

More recently, a new family of distributions has emerged that arise by integrating the Wigner distributions over the spatial position of the parton [3-3]. These structure functions, known as transverse momentum distributions (TMDs) hold information on the quark/gluon intrinsic motion in a nucleon, and on the correlations between the transverse momentum of the quark, and the quark/nucleon spins. They offer a unique opportunity for a momentum tomography of the nucleon, complementary to the spatial tomography of GPDs. TMDs can be measured in Semi-Inclusive Deep Inelastic Scattering (SIDIS), in which the nucleon is no longer intact and one of the outgoing hadrons is detected. We next describe Spatial and Momentum Tomography in more detail.

### Spatial Tomography of the Nucleon

Spatial densities (form factors) and longitudinal momentum densities (parton distributions) have in the past encapsulated our knowledge of the structure of the nucleon. GPDs have revolutionized how to characterize nucleon structure. GPDs impart a unified description of the response of the nucleon to scattering processes in which a short–distance probe interacts with a single quark in the nucleon. They describe the correlation between the spatial distribution of the quarks and its longitudinal momentum fraction, that is how the spatial shape of the nucleon changes when probing quarks of different wavelengths.

It is now recognized that DVCS and DVMP are the most powerful processes for providing the necessary observables to perform the spatial tomography of the nucleon for each constituent flavor. Mapping the vector GPDs $H(x, \xi, t)$ (nucleon non spin-flip) and $E(x, \xi, t)$ (nucleon spin-flip) in their momentum fraction of the struck quark $x$, longitudinal momentum fraction transferred to the quark $2\xi$ and transverse momentum transferred to the nucleon $t$, offers access to a three-dimensional image of the nucleon (with two dimensions in transverse space and one in longitudinal momentum) and to the complex dynamics of the constituents. Furthermore, as a direct consequence of the space-momentum correlation there is a way to reach the contribution the orbital angular momentum of quarks makes to the nucleon's spin through the Ji sum rule [3-19],

$$J^q = \frac{1}{2}\int_{-1}^{+1} dx\, x\,[H^q(x,\xi,t) + E^q(x,\xi,t)] = \Delta\Sigma^q/2 + L^q,$$

where $\Delta\Sigma$ is the well measured total spin carried by the quarks in the nucleon, while $J^q$ and $L^q$ are the total and the orbital angular momentum carried by quarks of flavor $q$ respectively.

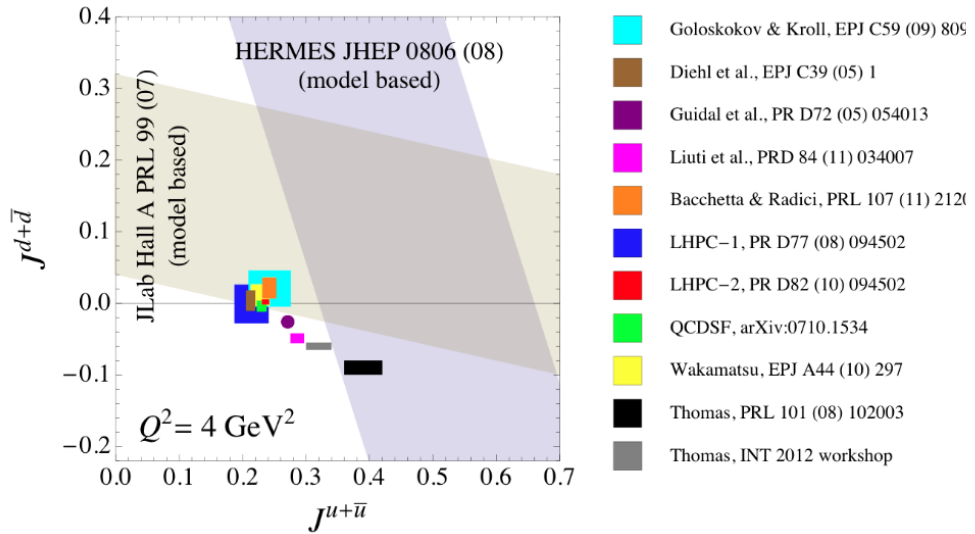

*Figure 3.2: Model dependent constraints on the total angular momentum carried by up quarks versus that carried by down quarks. The two overlapping bands are results from one DVCS experiment on a unpolarized neutron target at Jefferson Lab the other a HERMES DVCS experiment on a transversely polarized proton target. The colored boxes represent several calculations within lattice QCD and within quark models including a model dependent approach based on SIDIS data [3-5].*

The worldwide DVCS experimental program, including those at Jefferson Lab with a 6 GeV electron beam and HERMES with 27 GeV electron and positron beams, has given the first insight into the nucleon GPDs by allowing initial comparisons with models. These experiments have measured large asymmetries, in the 10 - 20 % range, and suggest an early approach to the scaling regime. Constraints on the total angular momentum of the $u$ and $d$ quarks from DVCS experiments are shown in *Figure 3.2*, together with calculations from Lattice QCD and QCD-inspired models.

Accessing GPDs requires the dedicated, long-term experimental effort that the deep exclusive reaction program of the 12 GeV upgrade provides. GPDs are not measured directly, but rather enter into different combinations and weighted integrals over $x$. To unravel them requires a diverse suite of experiments measuring a variety of observables, including cross sections, beam-spin asymmetries and target spin asymmetries for both longitudinal and transversely polarized targets. These measurements will be performed for a variety of channels, including both DVCS and DVMP for mesons having different isospins.

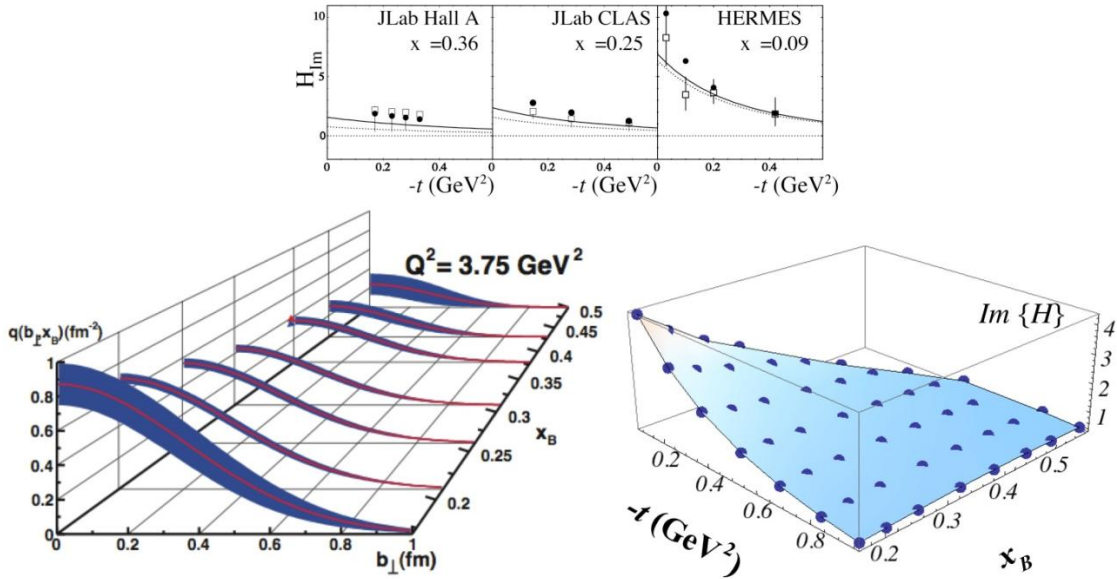

*Figure 3.3: Upper panel shows a determination of the Compton form factor Im{H} in the valence region as a function of t and x using the world data. The right bottom panel shows the impact of the 12 GeV measurements on this determination [3-65]. Note that both GPD H and E are needed to determine the angular momentum of the valence quarks. The left bottom panel is the nucleon transverse profile in terms of the impact parameter of a quark relative to the center of the nucleon at a fixed $Q^2$ and different x values. Note the narrowing of the profile as x increases [3-66]; the blue band is the projected error bar centered on the model GPD calculation.*

The increased energy of the electron beam to 12 GeV offers not only a reach of momentum transfer allowing for the leading order GPD formalism to be applicable, but also provides the highest polarized luminosity for precision measurements of key polarization observables crucial in these studies. A suite of approved DVCS experiments planned in Hall B [3-6, 3-7] with CLAS 12 and Hall A [3-8] will provide the necessary high precision data for different channels and reactions over a wide kinematical range. These data will be critical in the extraction of GPDs parametrizations, while constraints from dispersion-relation techniques and from the lattice calculations of the moments of GPDs, will minimize the model-dependence in those parametrizations. We show in *Figure 3.3* the projected impact of the 12 GeV data on our knowledge of the GPD *H* together with the current status of the 6 GeV data.

There is an equally ambitious program of experiments involving DVMP, which are able to access GPDs, or combinations thereof, that are inaccessible to DVCS. In Hall B, in parallel with the DVCS measurements of cross sections, structure functions and beam spin asymmetries both for vector mesons and pseudoscalar mesons will be explored over the largest phase space ever probed in the valence regime [3-50]. Experiments in Hall C will focus on L/T separation for pion electroproduction [3-52], and for the first time make precision measurements of $K^+$ cross sections adding strangeness information to the DVMP program [3-53].

## Momentum Tomography of the Nucleon

In the past 10 years, pioneering experiments at DESY (HERMES), CERN (COMPASS) and Jefferson Lab performing semi-inclusive deep inelastic lepton scattering (SIDIS) with polarized lepton beams and polarized targets (proton, deuteron and $^3$He) have offered a first glimpse of the effects of transverse motion of quarks and the way this is correlated with either their own spin or that of the nucleon.

In these experiments, a leading high-momentum hadron is detected in coincidence with the scattered lepton. It was realized rather quickly that encoded in the measured observables of the SIDIS process are TMDs. The TMDs are functions of the intrinsic longitudinal momentum fraction $x_B$ and the transverse momentum $k_T$ carried by the partons and they emerge naturally in the unified framework of the Wigner distributions once the position information is integrated out, *Figure 3.1* Accessing the transverse momentum of partons in the nucleon presents a special opportunity for a deeper view of the internal dynamics of the nucleon by tackling the relation between the orbital motion of quarks and gluons, their spin and the spin of the proton.

The general expression of a SIDIS cross section at leading twist is a factorized convolution of a TMD of an initial quark, an elementary photon-quark cross section, and a fragmentation function describing the hadronization of the struck quark into a hadron, for example into a pion or a kaon. The hadron resulting from the fragmentation of a scattered quark retains a memory of the original transverse motion of the quark, and thereby presents new information about the transverse momentum dependence of the quark within the nucleon. By performing precision SIDIS measurements of a variety of spin-dependent cross sections, or asymmetries, and by detecting different hadron species, it is possible to reveal both the flavor-dependent initial-quark momentum distributions, and the scattered-quark fragmentation functions, and to learn about dynamical correlations arising from the orbital motion and spin of the quarks and/or that of the nucleon.

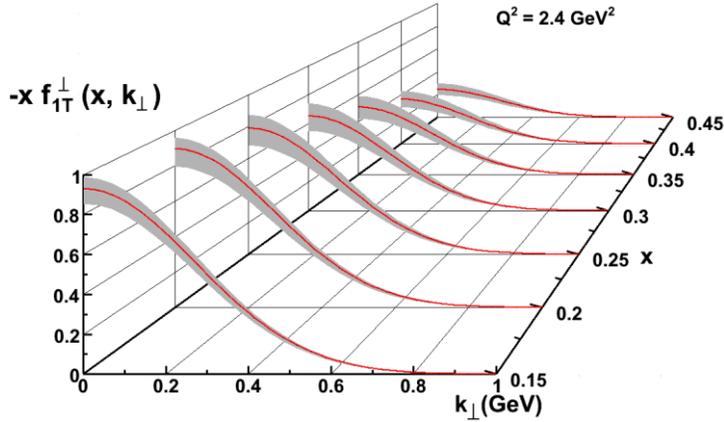

*Figure 3.4 The Sivers function for the up quark as a function of $k_\perp$ at different values of x as determined by analysis of JLab 12 pseudo data generated for $^3$He target. The central line is the model profile of [3-35]; real Jefferson Lab 12 GeV data will eventually reveal the actual shape of the distribution. The error bands have been projected about the model profile.*

Among the eight possible TMD functions, two have been a particular focus of theoretical and experimental studies, since they are responsible for large observed single spin asymmetries in the aforementioned SIDIS experiments. They are known as the Sivers function $f_{1T}^\perp$ and the Boer-Mulders function $h_1^\perp$. Both of these functions are related to the imaginary part of the interference of wave functions having non-zero orbital angular momentum. They describe unpolarized quarks in a transversely polarized nucleon, and transversely polarized quarks in an unpolarized nucleon, respectively. The Sivers transverse-momentum distribution function has been recently used [3-5] to infer the GPD *E* in the collinear limit in order to estimate the angular momentum carried by quarks in the nucleon. The results of this study are shown in *Figure 3.2*; though they depend on a relation, inspired by spectator models, between the Sivers function and the GPD *E*, the consistency with constraints arising from DVCS measurements is encouraging. In SIDIS this function is also responsible for an effect known as "color lensing", which describes the overall color attraction between a struck quark on its way to becoming a hadron and the remnant system [3-9]. Finally, an important prediction of QCD that needs to be confirmed in experiment is that the Sivers function determined in SIDIS has the opposite sign to that measured in a Drell-Yan experiment.

The multi-dimensional phase space of SIDIS is complex, with interesting and unknown physics reflected in each kinematic phase space. The 12 GeV era at Jefferson Lab can move this field to a new level of sophistication thanks to the extraordinary statistical accuracy achievable and the extended kinematic reach. Each hall brings an essential element to the SIDIS campaign: Hall B with the large-acceptance CLAS12 spectrometer will provide multi-dimensional cross sections, azimuthal distributions and single-and double-spin asymmetries on both polarized and unpolarized neutron and deuteron targets, and on unpolarized nuclear targets. Hall A will provide much-needed neutron information through their world-leading polarized $^3$He target. Finally, Hall C will add precision cross sections and their ratios for both pions and kaons.

High statistical precision measurements of semi-inclusive pion and kaon production with multidimensional binning in momentum transfer $Q^2$, invariant mass of the unobserved system *W*, the final tagged-hadron energy fraction *z* and transverse momentum $P_T$, are essential for performing a model independent extraction of the TMDs. Such experiments have been proposed and approved for the 12 GeV upgrade. Future data from the corresponding experiments in Hall B with CLAS 12 [3-10, 3-14], in Hall A with Super-BigBite [3-15] and with SoLID [3-16, 3-18] complemented with precision SIDIS experiments in Hall C [3-51, 3-52, 3-53] will allow a far more precise determination of the Sivers function as shown in *Figure 3.4*, and thereby enabling a high resolution momentum tomography of the Sivers function illustrated in *Figure 3.5*. As a direct consequence, the orbital angular momentum carried by valence quarks will be finally resolved.

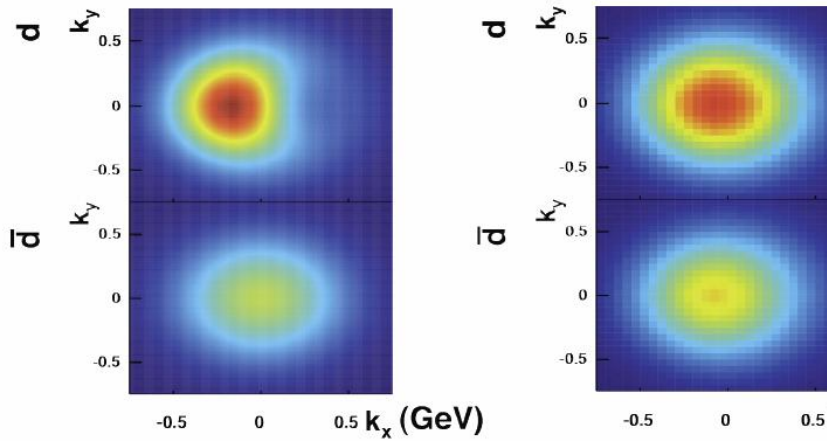

*Figure 3.5: Down quark momentum tomography of $f_{1T}^{\perp}$ at x = 0.1 for which a non zero value requires a non-zero orbital angular momentum of the quarks. The present available momentum tomography (right panel) does not resolve the dipole deformation while the achievable resolution with the 12 GeV upgrade (left panel) shows a clear deformation [3-65].*

The tensor charge is an important intrinsic property of the nucleon, similar to its axial charge or magnetic moment, and corresponds to the first moment in $x_B$ of the transversity distribution function $h_{IT}(x)$. It offers a benchmark test for the most modern Lattice QCD calculations. This distribution is accessible in SIDIS, through the well-known Collins' effect, by using transversely polarized targets. The tensor charge has been extracted using world data from Hermes, COMPASS and Jefferson Lab (6 GeV) with limited precision. It will be measured in Hall B using CLAS [3-14, 3-55] and in Hall A using SoLID [3-17, 3-18] and will be determined with much improved precision in the 12 GeV era.

## 3b    The Charge Radius and Flavor-Dependent Form Factors of the Proton

The electric and magnetic form factors of the nucleon describe the distribution of charge and currents, and are probed in elastic electron-nucleon scattering. At small values of the momentum $Q^2$ of the virtual-photon probe, they provide a description of proton size. Recent measurements of the proton radius from the Lamb shift in muonic hydrogen have revealed a 7σ discrepancy with the electron scattering measurements. This puzzle highlights how our limited knowledge of baryon structure also limits high precision tests of QED in atomic systems. The 12 GeV upgrade affords the prospect of addressing this puzzle in a timely and unique way with an electron scattering experiment at values of $Q^2$ down to $2 \times 10^{-4}$ GeV$^2$ [3-65].

At large value of $Q^2$, form factor measurements should reveal an approach to perturbatively dominated dynamics, and provide a further measure of the rôle of orbital angular momentum.

Recently the measured electric and magnetic form factors of the proton and neutron have been used to determine the flavor contributions to the Dirac and Pauli form factors. While there are a variety of QCD-inspired models that can describe the existing form-factor data for both the *up* quark and *down* quark in the proton at moderate $Q^2$, these models diverge dramatically at the larger values of $Q^2$ that will be probed with high precision by an array of approved 12 GeV experiments to measure the electromagnetic form factors of both the proton and the neutron [3-41, 3-42, 3-43, 3-44, 3-45, 3-46]. The unprecedented quality and breadth of these experiments will enable us to venture deep inside the nucleon to determine the distributions of charge and currents, and to unravel the flavor structure of the nucleon. The ability to discriminate pictures of QCD is shown in ***Figure 3.6.*** Further, these high-precision data will provide constraints on parameterizations of the GPDs discussed earlier.

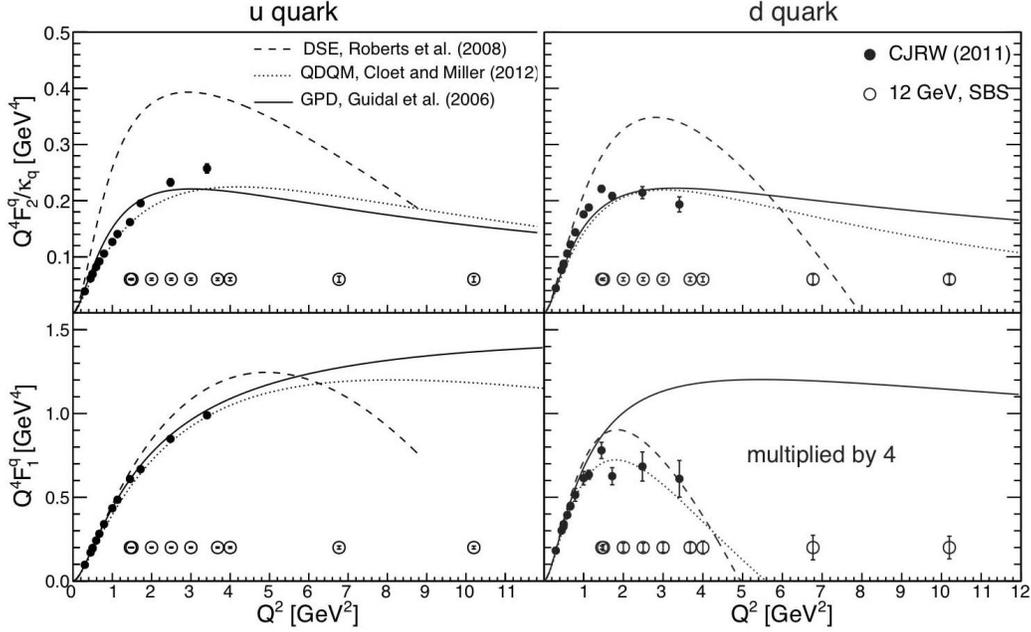

*Figure 3.6: The electromagnetic form factors are measured in elastic scattering, and characterized by the momentum $Q^2$ of the virtual photon. Flavor dependent Pauli (upper panel) and Dirac (lower panel) form factors for the up (left panel) and down (right panel) quark in the nucleon are shown as a function of $Q^2$. The present results (filled circles) [3-73] with the planned measurements (open circles) are compared to recent a Dyson-Schwinger equations (DSE) calculations for which the quark-diquark contribution at large momentum transfer behavior can be identified (dashed line) [3-74], a quark-diquark quark model calculation (dotted line) [3-75] and a GPD model [3-76] that provides a good representation of the existing data other than those of elastic form factors data [SOLID LINE].*

## 3c  *N\** Resonances and Transition Form Factors

The excitation of nucleon resonances has been a core component of the 6 GeV program. A unique data set for the observables of meson photo- and electro-production channels off nucleons measured with the CLAS detector has improved and extended the available information on N* photo- and electro-couplings, as well as on hadronic decay parameters [3-62]. *The Excited Baryon Analysis Center*, established at Jefferson Lab, has developed the tools for performing a dynamical coupled-channel analysis and applied those techniques to world pion- and photo-production data to provide unprecedented information on the spectrum of *N\** resonances. Recent lattice calculations suggest a spectrum of baryons at least as rich as that of the constituent quark model [3-63], and the presence of predominantly "hybrid" baryons with a gluonic excitation scale comparable to that expected for hybrid mesons [3-64] described in section 2a. These developments reinforce the need for a vigorous experimental program in the physics of nucleon excitations.

Understanding the structure of the nucleon is not limited to mapping out the charge or momentum distributions of its constituents in the ground states. As the paradigm of the Hydrogen atom has shown, an understanding of the structure of a bound state and of its excitation spectrum go hand in hand. Physics analyses [3-20, 3-21, 3-22, 3-23, 3-24, 3-25, 3-26, 3-27, 3-28] of the results for $\gamma NN^*$ electro-couplings that are available from the $N^*$ program with the current CLAS detector [3-29, 3-30, 3-31, 3-32] have elucidated the resonance structure at photon virtualities of $Q^2 < 5.0$ GeV$^2$, and revealed the important role of pionic degrees of freedom.

The CLAS12 detector in Hall B will be a unique facility worldwide to determine transition $\gamma NN^*$ electrocouplings of all prominent excited nucleon states in the still almost unexplored region of $Q^2$ from 5 to 12 GeV$^2$ [3-38]. Such studies explore the transition to the perturbative regime of QCD. When combined with appropriate theoretical calculations, this transition region exposes the strong coupling dynamics of QCD where both the running of the coupling and of the light quark masses influence the behavior of form factors. Furthermore, the experimental information will open access to the parton distributions in an excited nucleon, and enable the concept of GPDs to be applied to the transition of a nucleon to its excited state in which a quark is taken out of the nucleon and inserted into its excitation.

## 3d    The charged pion form factor

The pion is the lightest quark system in nature. Due to its simple quark-antiquark valence structure, the pion lends itself as a benchmark for QCD calculations, and to revealing its underlying dynamics. In QCD, the mass of quarks is dynamically generated, and the quark's effective mass depends on its momentum. Changing the spatial resolution with which one probes the pion effectively probes this behavior. As a result, the measurement of the electromagnetic form factor of the pion is the best hope for gauging our understanding of the expected transition from the large-distance scales with confinement-dominated dynamics at modest $Q^2$ to the short-distance scales with perturbative-dominated dynamics at high $Q^2$. This measurement is very timely given the experimental and theoretical controversy over the large $Q^2$ results for the $\pi\gamma^*\gamma$ form factor [3-33, 3-39] that has refocused attention on the need to understand the distribution of momentum between the valence quark and antiquark. With the 12 GeV Upgrade, the pion form factor can be accurately mapped in Hall C up to momentum transfers of 6 GeV$^2$ [3-40], as illustrated in *Figure 3.7*.

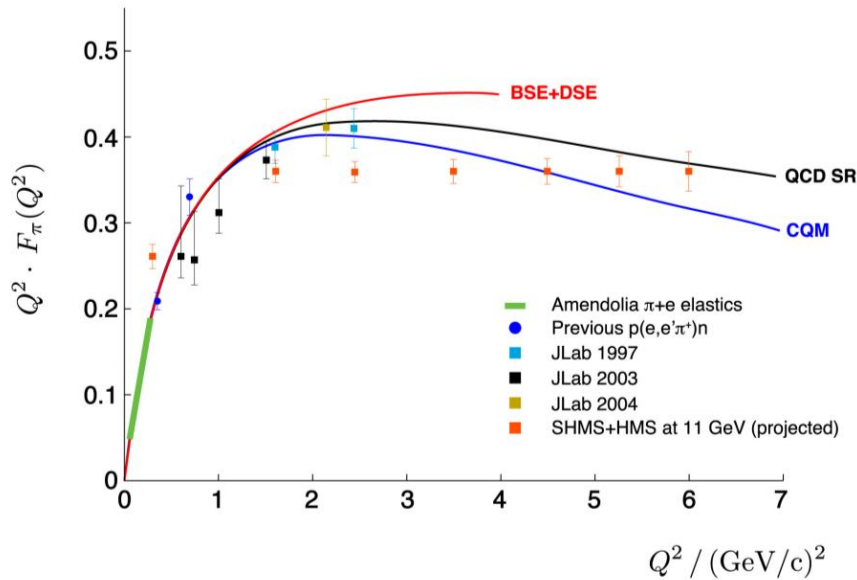

*Figure 3.7: The red points show projected data for the pion form factor after the 12 GeV upgrade [3-40]. Also shown is our current knowledge of the pion form factor, together with the different theoretical expectations as the red [3-77], black [3-78] and blue [3-79] lines.*

## 3e    Valence Structure of the Nucleon

One of the key properties of the nucleon is the structure of its valence quark distributions. Valence quarks are the irreducible kernel of each hadron, responsible for its charge, baryon number and other macroscopic quantum numbers. While deep inelastic scattering and other experiments have provided a detailed map of the nucleon's quark distributions at $x < 0.5$, there has there never been an experimental facility capable of accurately measuring the cross sections throughout the "deep valence region" ($x > 0.5$), where the three basic valence quarks of the proton and neutron predominate. Thus the large-$x$ region provides an ideal theater in which to explore the dynamics of the quarks in a nucleon. Moreover, an accurate knowledge of the distributions in the deep valence region has an important impact on hadronic cross sections at the small values of $x$ and large values of $Q^2$ relevant for the LHC, and can affect searches for new physics beyond the Standard Model [3-47, 3-48].

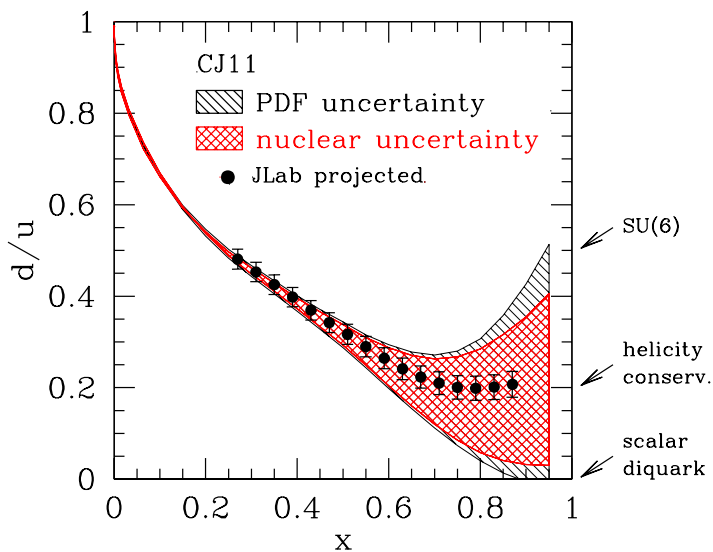

*Figure 3.8: The points show projected data for the ratio of the d-quark and u-quark distributions in the proton from experiments at Jefferson Lab 12 GeV Upgrade [3-69, 3-70]. The red band shows the uncertainties in the current CJ11 parton distributions arising from theoretical uncertainties on nuclear corrections to inclusive DIS on deuteron targets, while the grey band denotes those arising from the experimental errors on the fitted data.*

**Flavor structure at large-$x$**  The ratio of the *down* to *up* quarks in a proton, *d(x)/u(x)*, is a particularly sensitive measure of models of the nucleon, ranging from zero in models in which a scalar diquark dominates to one half in those with spin-flavor symmetry. Determining this ratio requires the measurement of both the neutron and the proton structure functions; obtaining the former is impeded by the lack of a free neutron target and the consequent need to disentangle the nuclear effects. The BoNuS experiment [3-34] at 6 GeV has demonstrated how the observation of the spectator proton in scattering from a deuterium target provides a powerful constraint on these nuclear corrections that will be exploited in the deep valence region at 12 GeV. A dedicated theoretical effort to quantify these effects has been established and the implications are shown in *Figure 3.8*. Simultaneous measurements will be made on $^3$He and $^3$H targets, mirror nuclei in which the nuclear effects are largely common [3-49]. What is more, the proposed experimental program of parity-violating deep inelastic scattering with the SoLID spectrometer includes measurements with a proton target; the comparable size of *u* and *d*-quark weak charges will enable *d/u* ratio measurements with comparable precision in the deep valence region without any nuclear corrections. Together, these programs will enable the ratio *d(x)/u(x)* to be extracted with unprecedented precision, as we show in *Figure 3.8* [3-69, 3-70, 3-71, 3-72].

**Spin Structure in the valence regime.** The uncertainties that characterize our present knowledge of the unpolarized distributions at large *x* are even more apparent in the polarized distributions that describe how spin is apportioned amongst the valence quarks, where once again the lack of a free neutron target serves to mask the valence behavior of the *down* quarks in particular. Whilst most dynamical models predict that the valence quark carrying most of the momentum at large *x* should also carry most of the spin, implying $\Delta d/d \rightarrow 1$, current Jefferson Lab data and analyses of world DIS data suggest the ratio remains negative. The data at 12 GeV obtained from measurements of the virtual photon-nucleon asymmetry in inclusive and semi-inclusive DIS [3-55, 3-56, 3-57] will resolve these contrasting descriptions, as shown in *Figure 3.9 left panel*.

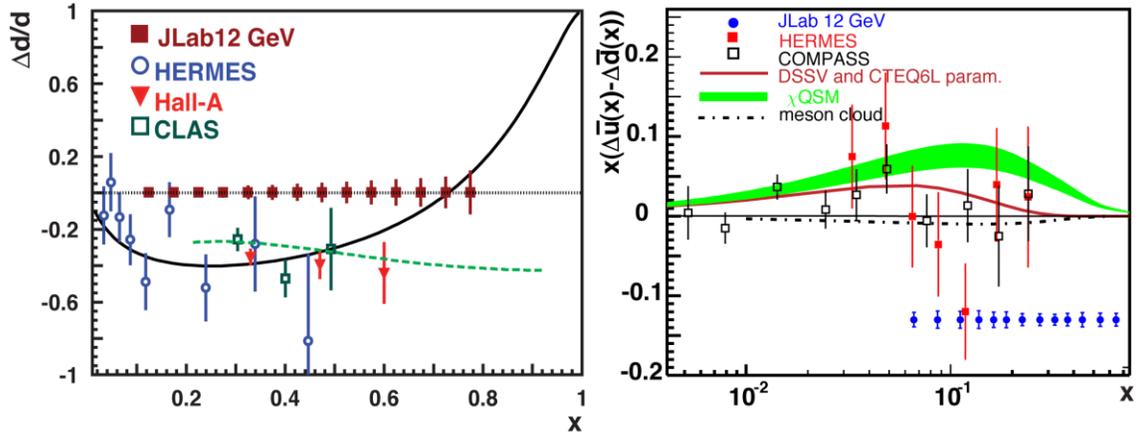

*Figure 3.9: Left panel: The blue points show current data for the distribution $\Delta d/d$. The red points show anticipated data from CLAS at 12 GeV. The yellow and purple lines show the expectation from perturbative QCD and in a constituent quark model respectively. Right panel: The red solid and black open squares show the present world data on the polarized sea helicity difference between the up and down quarks $x(\Delta \bar{u} - \Delta \bar{d})$ compared to various theoretical estimates. The blue points show anticipated data at the Jefferson Lab 12 GeV Upgrade [3-58].*

Although the contribution of sea quarks in the valence quark region is expected to be small, their helicity distributions at large $x_B$ are integral to our understanding of nucleon structure, yet are poorly known. Theoretical expectations from quark model calculations suggest that in the valence region a difference exists between the helicity carried by the sea down quarks compared to the sea up quarks. The extent of this difference will be clearly resolved with the 12 GeV data [3-58] as displayed in *Figure 3.9 right panel*.

Higher moments of spin structure functions offer a unique opportunity to connect observations with nucleon matrix elements of product of operators that are calculable using Lattice QCD. A linear combination of the spin observables $g_1$ and $g_2$, namely $(2g_1 + 3g_2)$, is of particular interest because it provides clean information on quark-gluon correlations. Its second moment, known as $d_2$, represents the restoring color force a struck quark experiences due to the remnant system in a deep inelastic electron scattering right at the point it is imparted momentum by the virtual photon. Measurements of such a quantity performed on the nucleon at 6 GeV [3-59, 3-60] and those planned at 12 GeV [3-61] will give us a glimpse of the size of the quark-gluon correlations. These measurements along with those of the Sivers function mentioned earlier provide a complementary input for our understanding of the struck quark propagation in the presence of a remnant system and the effect of color lensing.

# 4  QCD and Nuclei

Lying at the core of every atom is the nucleus with over 99.99% of the atom's mass but only $10^{-14}$ of its volume. It is a unique many-body system that can be understood quantitatively at low energies and momentum transfers as an assembly of protons and neutrons bound by an effective nuclear force. The dominant two-body *NN* force at "long range" of a few fermis is historically understood as mediated by the exchange of a single pion. This attractive force is balanced by a repulsive force at shorter distances.

But the nuclear force is the residue of the much stronger, non-perturbative QCD force between quarks, and the fundamental nature of the short distance repulsion is not well understood in either picture. Thus the delicate interplay between attraction and repulsion that enables the existence of atomic nuclei and therefore chemical elements is a vital topic for current research. After many decades of study, the basic mystery of the origin of nuclei still remains. How do nuclei emerge from QCD? Do any 'fingerprints' of QCD remain in the structure of nuclei?

The Jefferson Lab 12 GeV Upgrade will both study the QCD structure of nuclei and use the nucleus as a laboratory to study QCD. It will investigate a number of the most fundamental questions in modern nuclear physics:

- What is the nature of the nucleon-nucleon (*NN*) relative wave function at short distances? Can this be described in terms of nucleons and mesons, or are quarks and gluons necessary?

- The nuclear environment is known to modify the quark-gluon structure of bound nucleons. What is the nature of this modification and how is it related to the short-distance *NN* wave function?

- How thick is the neutron skin in heavy nuclei? What are the implications for neutron stars?

In addition, nature provides in the atomic nucleus a perfect laboratory to examine the underlying non-Abelian degrees of freedom of QCD, such as its elusive color charge. The study of how nature builds a fully dressed 1 GeV nucleon out of three current quarks with a total mass of 20 MeV has been discussed in section 3. However, the nuclear environment provides a new arena for studying this topic. By striking a single quark and observing the hadrons that emerge, we can study the physics of hadron formation, including the energy loss and cross section of the struck quark, the cross section of the postulated small size configurations, and the energy loss and cross section of the hadrons themselves.

## 4a  Electron scattering from nuclei

Inclusive electron scattering, $A(e, e')$, is a valuable tool for studying nuclei. By selecting specific kinematic conditions, especially the four-momentum and energy transfers, $Q^2$ and $\mathcal{V}$, one can focus on different aspects of the nucleus. Elastic scattering has been used to measure the nuclear charge distribution. Deep inelastic scattering at high $Q^2$ and $x_B \leq 0.7$ ($x_B = Q^2/2m\nu$, where $m$ is the nucleon mass) corresponds to scattering from a single quark, where $x_B$ is a measure of the quark's momentum fraction. ($x_B = 1$ corresponds to elastic scattering from a free proton and $x_B = 2$ corresponds to elastic scattering from a free deuteron.) Quasielastic scattering corresponds to scattering from a single bound nucleon. At large $x_B$ ($x_B > 1.5$) this is sensitive to high-momentum nucleons and to nucleon-nucleon short-range correlations (SRC) in the nucleus. We find nuclear modifications of these distributions in the comparison of both deep inelastic and quasielastic nuclear cross sections to those of deuterium. ***Figure 4.1*** displays the ratios of nuclear to deuterium cross sections (per nucleon) over a wide range of $x_B$ and nuclei. These ratios are very different from the naively expected value of unity.

## 4a.1   Short-range structure of nuclei

The short-range repulsive core is a critical component of the *NN* force. It leads to the saturation necessary for stable nuclei and causes substantial shell occupations at energies and momenta well above the Fermi sea. However its QCD origin remains unclear and little is known about the relative *NN* wave function at short distances.

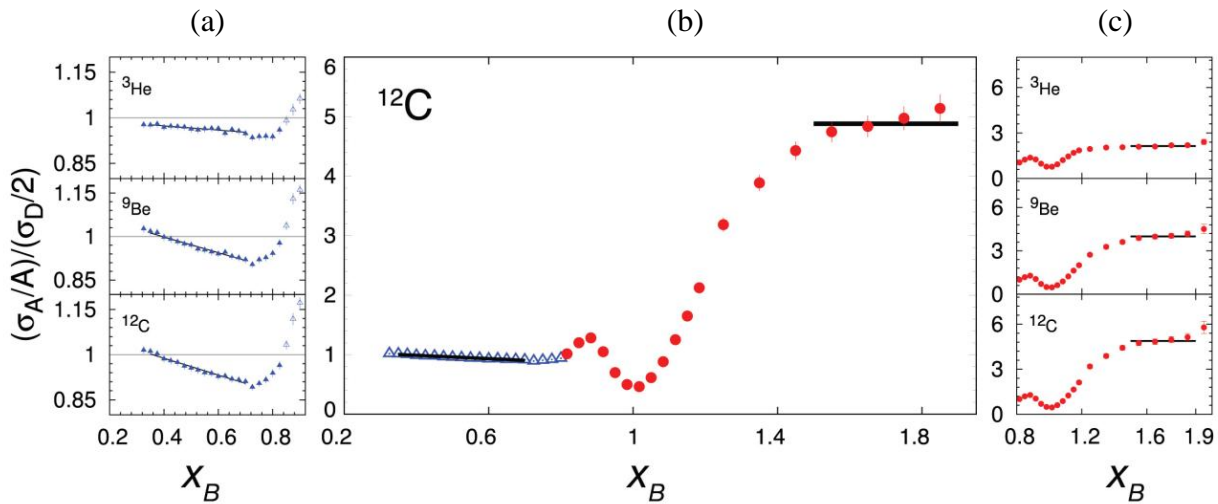

*Figure 4.1: The electron scattering cross section ratios (per-nucleon) of various nuclei to deuterium. (a) Ratios for $0.2 < x_B < 0.9$ [4-1]; (b) Ratios for $^{12}C$ for $0.3 < x_B < 2$ [4-1, 4-2]; (c) Ratios for $0.8 < x_B < 1.8$ [4-2].*

Nucleons acquire high-momentum via short-range interactions, rather than by the influence of the nuclear mean-field. Thus, for momenta above the Fermi sea, the shape of the momentum distribution should be universal. This is shown by the plateaus (regions of constant cross section ratio) observed at $x_B > 1.5$ (see ***Figure 4.1c***). The values of the ratio at these plateaus, $a_2(A/d)$, are closely related to the probability of finding a nucleon belonging to a short-range correlated pair in nucleus *A* relative to deuterium.

Experiments at 12 GeV will study the deuteron to better understand the simplest nuclear system [4-18]. Other experiments will extend the SRC studies described above to a much greater range of $x_B$, momentum transfers, and nuclei (including $^3$H and $^3$He) to study two and three nucleon correlations (including their isospin character), with higher nucleon momenta [4-19, 4-20]. The highest $Q^2$ data from the $x_B > 1$ measurements will probe the distribution of superfast quarks in nuclei, greatly extending our understanding of nucleons at short distances.

## 4a.2 Nucleon properties in nuclei

Another outstanding question is whether the nuclear medium alters the structure of bound nucleons and, if so, how? The neutron lifetime in nuclei is certainly different. The first evidence for nucleon structure modification was the EMC effect [4-3], in which deep-inelastic scattering from nuclear quarks is significantly different than from quarks inside a "free" nucleon (see *Figure 4.1a*). The per-nucleon DIS cross section ratio of nucleus $A$ to deuterium decreases approximately linearly for $0.3 < x_B < 0.7$. The slope of this ratio in this region increases with $A$ but does not scale simply with average nuclear density (see *Figure 4.2*). Despite a world-wide effort in experiment and theory, the origins of the EMC effect remain unclear. The 12 GeV experimental program will measure the EMC effect in a wide range of nuclei to help unravel the mystery [4-24, 4-25].

Recent phenomenological comparisons [4-4] (see *Figure 4.2*) show that the strength of the EMC effect in different nuclei is linearly related to the short range correlations scale factor, $a_2(A/d)$. This linear relation indicates, but does not prove, that the EMC effect is caused by *local* modifications of nucleon structure occurring when two would-be nucleons make a close encounter and briefly comprise a system of density high enough to be comparable to that of neutron stars. This relationship will be tested and refined at 12 GeV by a series of EMC and SRC experiments covering a wide range of nuclei [4-19, 4-20, 4-24, 4-25]. The deuteron experiment mentioned above [4-18] will increase our understanding of the two-nucleon system that we use as a baseline for both the EMC and SRC measurements.

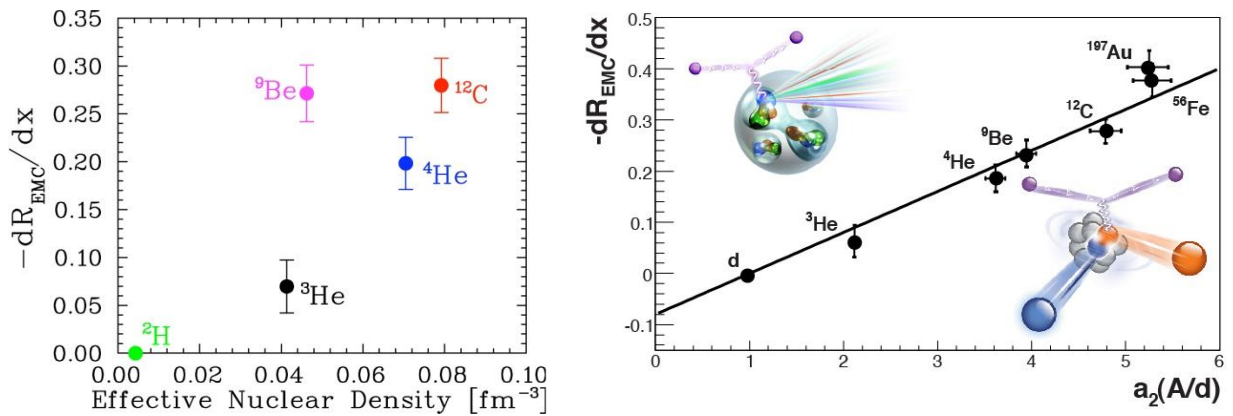

*Figure 4.2: (left) The strength of the EMC effect (the EMC slope) plotted versus the average nuclear density [4-1]. (right) The strength of the EMC effect plotted versus the SRC scale factors [4-4]. The drawing in the upper left shows deep inelastic electron scattering from a quark in a nucleon. The drawing in the lower right shows electron scattering from a correlated NN pair. (Figure credit: Anna Shneor).*

This linear relationship suggests a fertile field of research at 12 GeV: exploring the local nature of the EMC effect by measuring electron deep inelastic scattering in coincidence with low-energy spectator nucleons or target fragments. This technique has been pioneered at Jefferson Lab by detecting recoiled protons with low-momentum (relative to the nuclear Fermi momentum), that "tag" scattering events on nearly on-shell neutrons in a deuteron target [4-5] and detecting high-momentum recoil protons to study off-shell neutrons [4-6]. If the EMC effect is indeed caused by modification of nucleon structure occurring when two nucleons are at short range, then we expect the tagged nucleon structure functions (closely related to the tagged cross sections) to depend strongly on the initial parallel momentum of a struck nucleon, $P^z_{initial}$. Two deuteron-target experiments will explore this. The BoNuS12 experiment will use a recoil detector and the CLAS12 spectrometer to detect low-energy back-angle recoil protons in conjunction with high-energy electrons to tag deep inelastic scattering events occurring on low-momentum neutrons [4-7]. The LAD (E12-11-107) experiment will use a large-acceptance back-angle scintillator detector and the Hall C electron spectrometers to detect high-momentum back-angle recoil protons and neutrons to tag deep inelastic scattering events occurring on high-momentum nucleons [4-8]. ***Figure 4.3*** shows the expected change in $F^{eff}_{2p} / F_{2p}$, the ratio of the effective bound structure function to the free one, as a function of nucleon initial momentum. These ratios significantly differ from unity, and the large differences among models seen in ***Figure 4.3*** is indicative of how little is known.

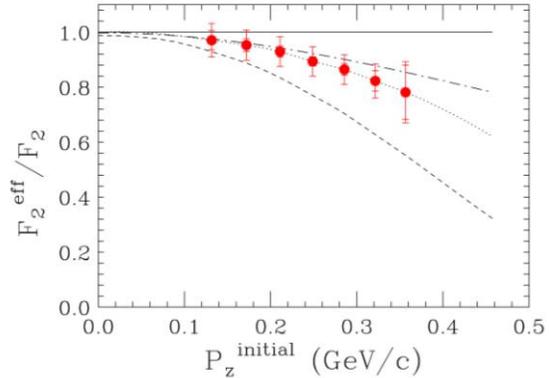

*Figure 4.3: Projected measurement uncertainties (statistical and systematic) of the dependence of the ratio of the bound to free neutron structure functions $F^{eff}_{2n}$ / $F_{2n}$ in deuterium as a function of recoil momentum $P^z_{intial}$ [4-8]. The lines show the different models of Ref. [4-9].*

A related experiment will measure the ratio of the transverse and longitudinal transferred polarizations in the $\vec{e}, e'\vec{p}$ reaction on the proton, deuteron and $^4$He [4-26]. For the free proton this ratio is proportional to $G_E/G_M$. The interpretation of this ratio is more complicated for bound protons. However, measuring this ratio on bound protons at higher precision over a much greater range of $Q^2$ and recoil momentum will greatly increase our understanding of nucleon modification in the nucleus

## 4b    Fundamental QCD processes in the nuclear arena

The 12 GeV Upgrade will provide unique opportunities for understanding how hadron-hadron interactions emerge from the underlying quark-gluon structure of QCD. The non-Abelian nature of QCD gives the gauge bosons — the gluons — color charge and causes them to strongly interact with each other. As already mentioned in section 3, the origin of quark confinement must reside in this feature. Here, we will use this confinement property as a tool: quarks, when struck violently by an 11 GeV electron, must turn into hadrons over a range comparable to the size of the nucleus. This will allow, for example, studies of quark propagation and hadronization in the nuclear medium and the onset of the phenomenon of color transparency in exclusive processes.

## 4b.1 Quark propagation through cold QCD matter: nuclear hadronization and transverse momentum broadening

Isolated quarks are experimentally inaccessible because of color confinement in hadrons. However, when an electron strikes a quark in deep inelastic scattering, the quark can be knocked out from its parent nucleon. As the quark propagates, it will interact with the nucleus and $q\bar{q}$ pairs will tunnel out of the vacuum. Eventually, fully formed hadrons emerge. By changing the energy and momentum transferred to the struck quark we can vary the distance over which the quark forms hadrons (hadronizes). In addition, choosing different targets allows us to vary the nuclear distance that the propagating quark traverses. At 12 GeV we will measure how the production of hadrons varies with kinematics and nuclear sizes to learn how the color field of the hadron is restored [4-17].

As an illustration, consider the transverse momentum distribution of the hadron emerging from larger nuclei. The propagating quark is expected to undergo multiple soft scatterings mediated by gluon emission. This causes a medium-induced energy loss, and a broadened transverse momentum distribution that should be measurable and which may exhibit exotic coherence phenomena. This will be a clear way to differentiate quark/color interactions from purely hadronic interactions.

The topics of color field restoration by gluon emission, quark-gluon correlations, and quark energy loss offer fundamental and interesting insights into the nature of QCD and confinement. Moreover, these are the basic ingredients required to understand the study of relativistic *AA* and *pA* collisions.

## 4b.2 Color transparency

The nature of hadronic interactions as the residue of the QCD force between quarks can be investigated by testing the predictions of "color transparency", one of the few direct manifestations of the color forces underlying nuclear physics [4-21]. Color transparency predicts that three quarks in a color singlet state, each of which (alone) would have interacted very strongly with nuclear matter, can form an object that passes undisturbed through the nucleus. A similar phenomenon occurs in QED, where a small-size $e^+e^-$ pair has a small electric dipole moment and hence a small cross section. In QCD, a small-size $q\bar{q}$ or $qqq$ system should have a small color dipole or tripole moment, and thus should have a small cross section due to cancelation of the color fields of the quarks. While $q\bar{q}$ color transparency is somewhat analogous to the QED example, color transparency in the $qqq$ case would be one of the rare demonstrations of the *SU(3)* nature of the underlying color degrees of freedom. However, we do expect to observe $q\bar{q}$ color transparency at lower momentum transfers than $qqq$ color transparency, since it is easier to produce a small-size two-quark system than a three-quark system.

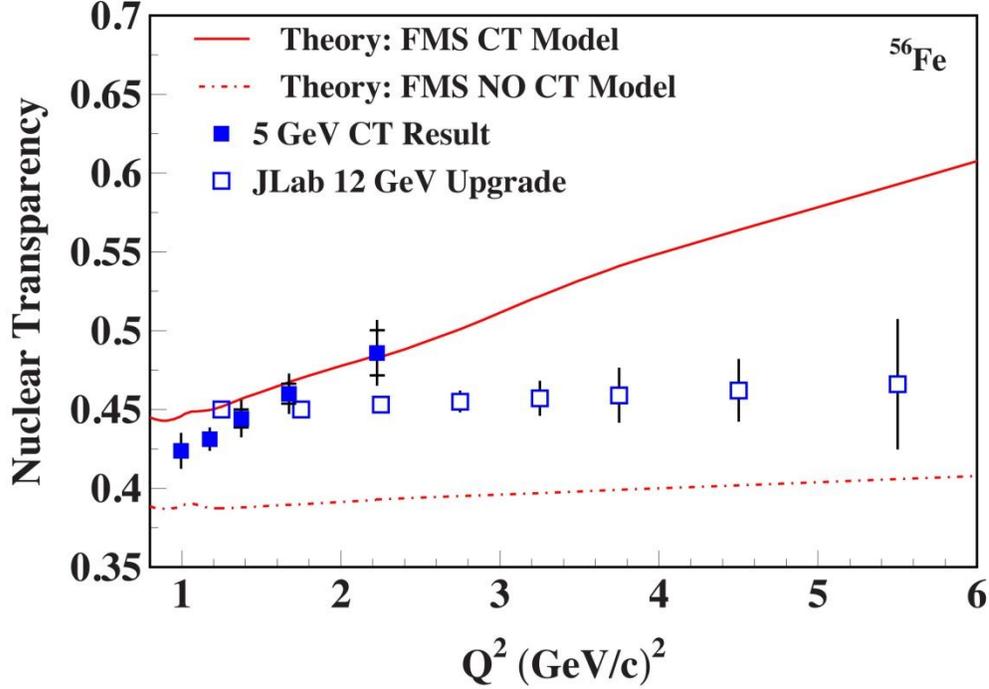

*Figure 4.4: Nuclear transparency, the per-nucleon ρ production cross section for iron relative to deuterium, as a function of $Q^2$ [4-10]. The curves show the predictions of Ref. [4-11] (red) with and without the effects of color transparency (CT). The open points show the expected precision for the Jefferson Lab 12 GeV measurement [4-12].*

Color transparency can be observed experimentally by measuring the number of particles produced on nuclear targets as a function of momentum transfer. The effects of color transparency have been observed for dijets [4-13, 4-14] produced coherently from nuclei by incident 500 GeV pions, and strong hints of color transparency at lower Jefferson Lab energies have been observed in pion [4-15] and rho meson production [4-10] (see ***Figure 4.4)***. However confirmation and testing the predictions for larger values of $Q^2$ are needed to firmly establish color transparency for pion [4-23] and rho [4-12] meson production. These definitive tests will be performed at 12 GeV.

## 4c    Parity Violation and the Neutron Skin

The iconic use of electron-nucleus scattering has been the determination of nuclear charge densities, which are dominated by protons. In contrast, our knowledge of neutron densities is much less precise because it comes primarily from hadron scattering experiments involving the non-perturbative strong interaction. However, the weak charge of the neutron is much larger than that of the proton. This means that the measurement of the parity violating asymmetry, $A_{\rm PV}$, in elastic scattering of polarized electrons provides a model-independent probe of neutron densities, free from most strong interaction uncertainties. The first measurement using a $^{208}$Pb target was recently carried out in Hall A [4-16]. The measurement corresponds to a difference between the radii of the neutron and proton distributions of about 0.3 fm (see ***Figure 4.5***). This is the first electroweak observation of the neutron skin in a heavy, neutron-rich nucleus. The improved Jefferson Lab 12 GeV Upgrade experiment should reduce the uncertainties significantly for $^{208}$Pb [4-22] and allow comparable measurements for $^{48}$Ca.

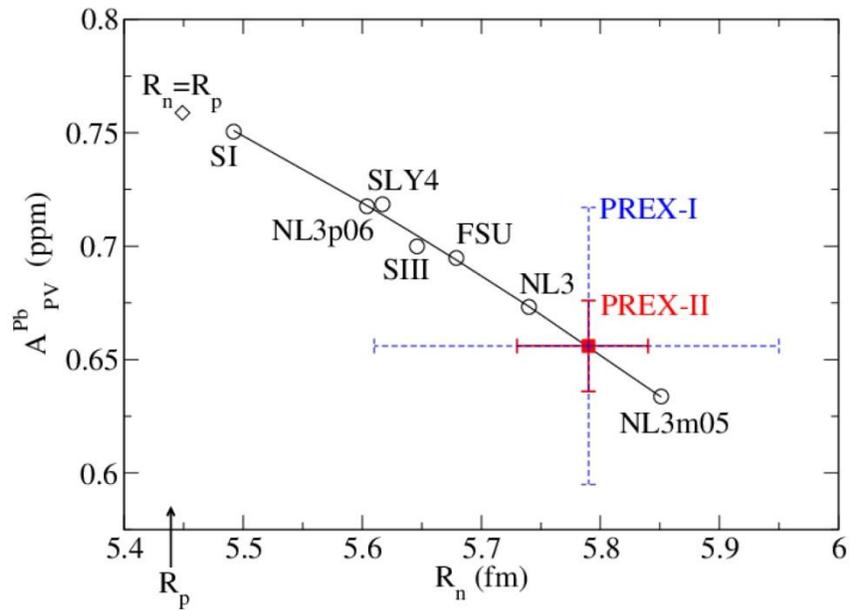

*Figure 4.5: The $^{208}$Pb neutron radius as determined by the parity-violating asymmetry [4-16]. The blue point and dashed error bars show the current results, the red point and solid error bars show the expected result of the 12 GeV experiment [4-22]. The diamond shows the result if the neutron and proton radii are the same and the open circles show the results of different models. The line is the result of a quadratic representation of the neutron radius as a function of the asymmetry.*

# 5   The Standard Model and Beyond

The Standard Model as a theory of the fundamental interactions is believed to be incomplete. It is likely part of a bigger theory with new particles and interactions that are expected to resolve many outstanding conceptual issues. The nature of new physics can be revealed at the low energy precision frontier accessed by Jefferson Lab, complementing information from the high energy frontier at the LHC. Importantly, a program of precision electroweak measurements and sensitive heavy photon searches will continue to be pursued at Jefferson Lab with substantial potential to discover and/or elucidate new physics.

## 5a   Parity-violating Electron Scattering

The dominant contribution to parity violation in electron scattering is from interference between neutral weak and electromagnetic amplitudes. The usual method for experimental studies of parity-violating electron scattering involves a longitudinally polarized electron beam incident on an unpolarized target. One periodically flips the helicity of the electron beam and measures the fractional change in the scattering rate, known as the parity- violating asymmetry $A_{PV}$:

$$A_{PV} = \frac{d\sigma_R - d\sigma_L}{d\sigma_R + d\sigma_L} \simeq \left[\frac{-G_F Q^2}{4\sqrt{2}\pi\alpha}\right] \times \mathcal{F}$$

Here, $d\sigma_L$ and $d\sigma_R$ are the differential cross sections for left and right helicity electrons, respectively. Since $A_{PV}$ measures the ratio of the weak and electromagnetic amplitudes, $\mathcal{F}$ characterizes the properties of the target and the neutral weak interaction (mediated by the $Z^0$ boson) and potentially contains clues to as yet undiscovered short distance dynamics.

For low $Q^2$, one generally finds the asymmetry to be quite small, parts per million, and so precise measurements are particularly challenging. A fundamental experimental issue is to control beam properties under helicity reversal in order to reduce systematic errors associated with beam fluctuations (such as slight position or angle shifts). CEBAF is an ideal facility for studies of parity violating electron scattering. Giant technical strides have been made over the past fifteen years, including a large increase in polarized luminosity, exquisite beam stability and improved absolute calibration of the longitudinal beam polarization. With the resulting ability to achieve part per billion precision and sub-1% normalization accuracy on $A_{PV}$ measurements, it has now become possible to probe for extensions of the Standard Model of electroweak interactions, which may for example involve the interactions of new massive ($M > M_Z$) bosons that couple to electrons and/or quarks, or other similar process [5-1].

In the Standard Model, the vector and axial-vector couplings of fermions to the $Z^0$ boson are precisely predicted. The vector couplings are functions of one free parameter: the weak mixing angle $\theta_W$. The most precise determinations of this parameter were obtained at the Z boson mass scale by $e^+e^-$ collider experiments. The two most precise of these measurements (with errors of 0.00029 and 0.00026) differ by ~ 3σ. Nevertheless, they are the main contributors to the world average value 0.23116 ± 0.00013 [5-5]. Radiative corrections associated with fermion and massive vector boson loops predict a "running" of this coupling to $\sin^2\theta_W = 0.2388$ at $Q^2 = 0$. *Figure 5.1* shows the Standard Model prediction ($\overline{MS}$ scheme) for $\sin^2\theta_W$ as a function of energy scale $\mu$. The three most precise measurements at energy scales below the Z mass are shown: parity violation in the $6S \to 7S$ transition of atomic $^{133}$Cs [5-2], Fermilab NuTeV neutrino deep-inelastic scattering cross-sections [5-3] and SLAC E158 $A_{PV}$ in electron-electron (Møller) scattering [5-4].

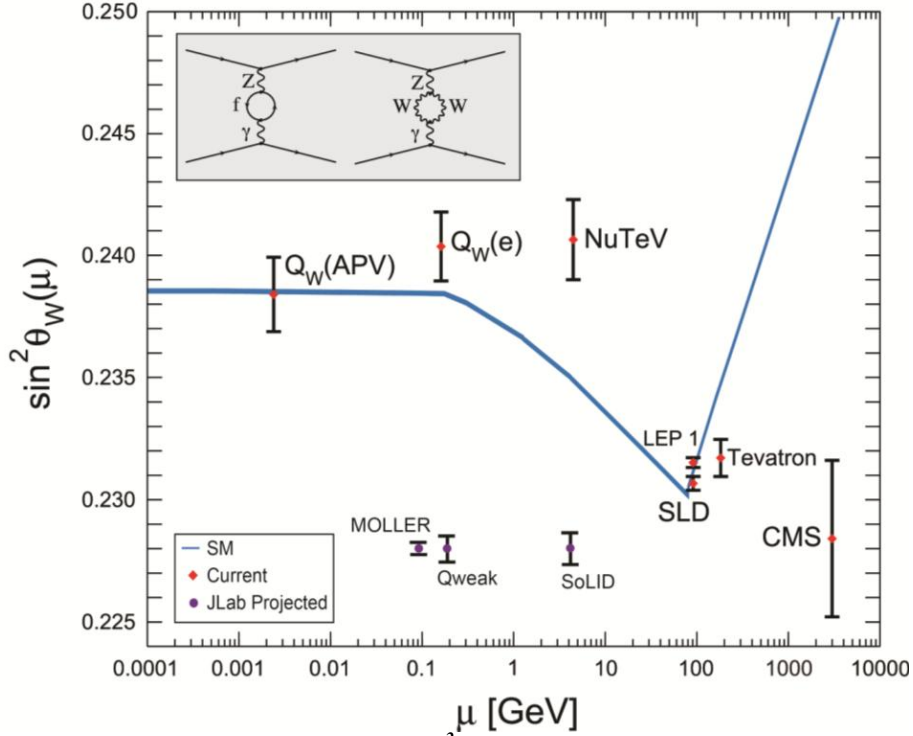

*Figure 5.1: The Standard Model prediction for $\sin^2\theta_W$ as a function of the energy scale $\mu$. One class of Feynman diagrams responsible for its "running" is also shown, along with published measurements (red) [5-5] and projected precision (shown at arbitrary y-axis values) for current and proposed Jefferson Lab projects.*

The effects of new high-energy phenomena on low energy measurements can often be characterized by new four-fermion contact interaction amplitudes that would contribute to changes in the overall parity-violating interaction with electrons. The contributions of such new amplitudes between two fermion species $i$ and $j$ are characterized by coupling constants $g_{ij}$ for the combination of four left- and right-handed fermion chirality projections, and a mass scale $\Lambda$. The goal of low energy neutral current measurements is to reach a sensitivity on contact interactions to access $\Lambda/g_{ij} \geq 1$ TeV for as many different initial and final state fermions as possible, similar sensitivity to that of the highest energy colliders.

It is useful to parameterize the parity-violating component of the interaction between electrons and target quarks or electrons in terms of phenomenological couplings $C_{ij}$:

$$\mathcal{L}^{PV} = \frac{G_F}{\sqrt{2}} \left[ \bar{e}\gamma^\mu\gamma_5 e \left( C_{1u}\bar{u}\gamma_\mu u + C_{1d}\bar{d}\gamma_\mu d \right) + \bar{e}\gamma^\mu e \left( C_{2u}\bar{u}\gamma_\mu\gamma_5 u + C_{2d}d\gamma_\mu\gamma_5 d \right) + C_{ee}\bar{e}\gamma^\mu\gamma_5 e \left( \bar{e}\gamma_\mu e \right) \right]$$

Here $C_{1j}$ ($C_{2j}$) depends on the vector (axial-vector) coupling to the $j^{th}$ quark and $C_{ee}$ is the four-electron coupling. The atomic Cs measurement constrained one combination ($C_{1u} + C_{1d}$) precisely, while E158 made the first ever measurement of $C_{ee}$.

*In Figure 5.1* the effects of new contact interactions would manifest themselves as lack of agreement with the Standard Model prediction for the extracted value of $\theta_W$ at the energy scale of each measurement. The three published low energy measurements shown, from atomic Cs, E158, and NuTeV, have set competitive limits on TeV-scale contact interactions. It should be noted that the nuclear corrections for the NuTeV result are still a subject of substantial discussion. The projected results for $\sin^2\theta_W$ for Jefferson Lab projects are shown at the correct energy scale, but arbitrary values of $\sin^2\theta_W$, to illustrate the expected experimental errors.

The *Qweak* experiment has completed data collection to measure $A_{PV}$ in elastic electron-proton scattering at low $Q^2$ in Hall C [5-6], which will provide a very precise constraint of a different linear combination of coupling constants i.e. $2C_{1u}+C_{1d}$. The PVDIS experiment [5-7] has completed data collection to measure $A_{PV}$ in deep inelastic scattering off deuterium. In addition, there are presently two new proposals to measure $A_{PV}$ at the upgraded CEBAF. One proposal involves the construction of a novel dedicated toroidal spectrometer to study parity-violating Møller scattering [5-8]. The other proposal would use a solenoidal magnetic spectrometer system (SoLID) to study parity-violating deep inelastic scattering [5-9]. Both experiments will require construction of substantial new experimental equipment (beyond the scope of the present upgrade project).

## 5a.1 Parity Violation in Electron-Electron (Møller) Scattering

Møller scattering is a purely leptonic process and a particularly clean way to test the Standard Model with high precision. The E158 experiment at SLAC [5-4] yielded the first measurement of $A_{PV}$ in this process resulting in one of the two best low energy measurements of $\sin^2\theta_W$ at low $Q^2$. The 12 GeV upgrade provides the unique opportunity to improve on the measurement of $A_{PV}$ in Møller scattering with a precision approaching 2%, a factor of 5 improvement over the E158 result. A new experiment called MOLLER [5-8] has been proposed to achieve a constraint on $\delta(C_{ee}) = 0.0011$, which would have a reach in sensitivity for new four-electron contact interaction amplitudes as small as $1.5\times10^{-3}\times G_F$. This corresponds to a sensitivity of $\Lambda/g = 7.5$ TeV, providing *the* most sensitive probe of new flavor and *CP*-conserving neutral current interactions in the leptonic sector until the advent of a linear collider or a neutrino factory.

MOLLER will achieve $\delta(\sin^2\theta_W)$ better than ±0.0003, precision comparable to the best high-energy measurements mentioned earlier, or that of the *W* mass. Should the LHC find evidence for new physics, it becomes particularly important to test the agreement between the directly measured value of $m_H$, and that inferred from the measurements of fundamental electroweak parameters $\sin^2\theta_W$ and $m_W$, and the mass of the top quark. *Figure 5.2* shows $m_H$ constraints from $\sin^2\theta_{eff}$ (as extracted from leptonic couplings evolved to $Q \sim M_Z$) from the best high and low energy measurements. The MOLLER precision is particularly relevant in light of the tension between the two most precise published measurements. Also shown in the figure are the constraints (green bands) on the Higgs mass taking into account the most recent data at the LHC; the narrow band at ~125 GeV is the allowed mass range of the recently announced discovery of a new resonance whose decay characteristics are largely consistent with those expected for a Standard Model Higgs boson [5-60, 5-61]. There is the intriguing possibility that new TeV-scale dynamics invalidates the $m_H-\sin^2\theta_W$ correlation for some or all of the measurements shown in the figure. Additionally, MOLLER would be the first low $Q^2$ measurement capable of providing a meaningful $m_H$ constraint. Given the current status of low $Q^2$ measurements (black points), a large range of $\sin^2\theta_w$ value outcomes remain viable.

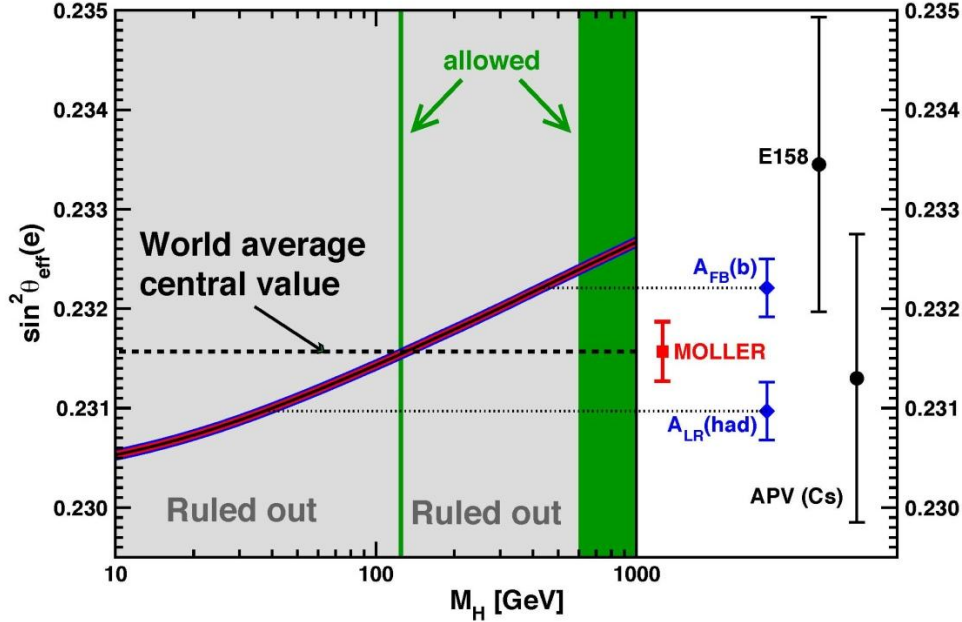

*Figure 5.2: Indirect $m_H$ constraints from the best high $Q^2$ (blue) and low $Q^2$ (black) $\sin^2\theta_w$ measurements, evolved to the same energy scale (courtesy Jens Erler). Green bands: allowed $m_H$ regions including LHC data released as of July 2012. Red curve: theory prediction for $\sin^2\theta_w$ vs $m_H$. Black dashed line: world average $\sin^2\theta_w$ from all electroweak asymmetry data. Red: proposed MOLLER precision (placed at an arbitrary y-axis value).*

MOLLER would utilize a 75 µA beam of 11 GeV electrons incident on a liquid hydrogen target. A normally conducting toroidal magnetic spectrometer selects the Møller scattered electrons in the angular range 5-18 mr. The predicted parity-violating asymmetry is $A_{PV}$ = 35.6 ppb, and the goal is to make a measurement with a projected statistical uncertainty of 0.8 ppb.

## 5a.2 Parity Violation in Deep Inelastic Scattering

The measurement of $A_{PV}$ in $e$-$^2$H deep inelastic scattering was the first experimental demonstration that the neutral current violated parity as expected in the Standard Model [5-10]. The most promising method for accurate determinations of the axial-quark couplings is to carry out more precise measurements of the same quantity. The use of a deuterium target minimizes the uncertainties due to quark distribution functions and the measurement is sensitive to the couplings $(2C_{1u} - C_{1d})$ and $(2C_{2u} - C_{2d})$, where the relative weight depends on the details of the kinematics, but is different from the combination accessed by the *Qweak* experiment. At the level of precision envisioned, small corrections to $A_{PV}$ also arise from strange quark distribution functions, possible charge symmetry violation, and higher twist effects. The optimum strategy would achieve simultaneous $A_{PV}$ measurements in narrow $(x_B, Q^2)$ bins to provide precision measurements of the $C_{iq}$ combinations, while simultaneously providing new insights into nucleon structure in the valence region.

*Figure 5.3* shows the projected improvements in the determination of $C_{iq}$'s; this is particularly remarkable for the axial-vector quark couplings. The proposed SoLID [5-8] project will improve the determination of $(2C_{2u}-C_{2d})$ by a factor of 6 to 8 over the 6 GeV initial measurement. The Jefferson Lab program will thus improve the current world data by more than a factor of 30. Quite apart from unambiguously establishing a non-zero value for a $C_{2q}$ combination for the first time, the measurement has complementary sensitivity to new TeV-scale physics. One example [5-11] is due to the sensitivity via radiative corrections to the so-called leptophobic Z′. Such a Z′ couples only to quarks and therefore could have so far escaped detection in semileptonic processes in both fixed target experiments as well as at colliders.

The experimental design involves a superconducting solenoidal spectrometer with collimators to select kinematics of the scattered electrons in the range $22° < \theta < 35°, x_B > 0.55$, and $W > 2$ GeV. High rate Gas Electron Multiplier detectors will track the scattered electrons, and a threshold gas Cerenkov detector plus electromagnetic calorimeter will enable pion rejection.

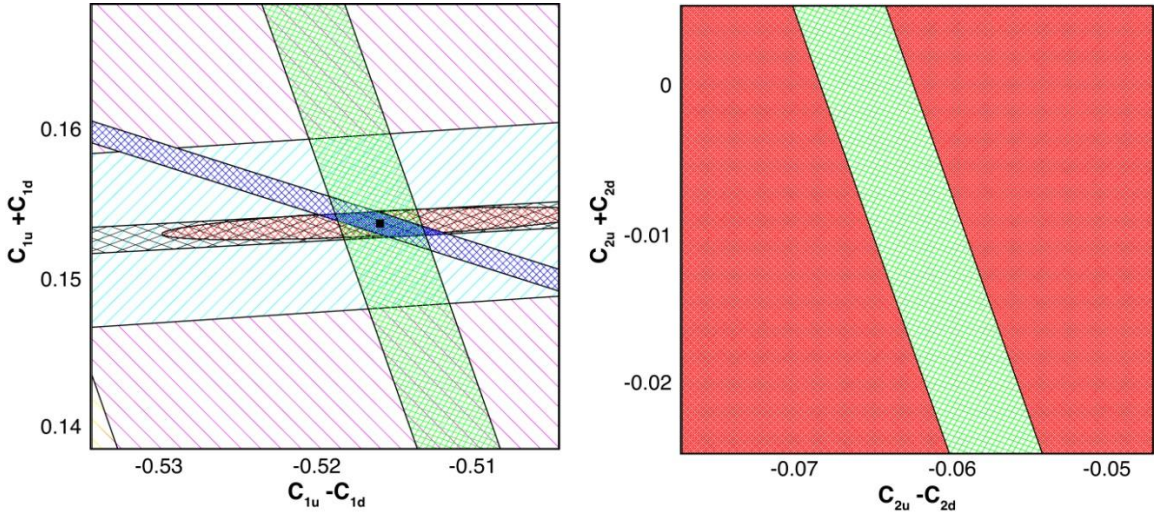

*Figure 5.3: Constraints (±1σ error bands) on electron-quark coupling constants from parity-violation experiments. The black crossed band presents the $A_{PV}$ results, the blue band - the expected $Q_{weak}$ result the red ellipse is a fit of existing data from the Particle Data Group (PDG), while the green bands shows the projected constraints from SOLID. The entire region of the right plot is unconstrained.*

## 5b    New Searches for Heavy Photons

Heavy photons, called *A′*s, are new hypothesized massive vector bosons that have a small coupling to electrically charged matter, including electrons. The existence of an *A′* is theoretically natural and could explain the discrepancy between the measured and observed anomalous magnetic moment of the muon [5-59] and several intriguing dark matter-related anomalies. New electron fixed-target experiments proposed at Jefferson Lab, with its high-quality and high-luminosity electron beams, present a unique and powerful probe for *A′*s. These experiments include the *A′ Experiment* (APEX), *the Heavy Photon Search* (HPS), and *Detecting A Resonance Kinematically with Electrons Incident on a Gaseous Hydrogen Target* (Dark Light). We briefly review the theory and motivation for heavy photons, before discussing the rich experimental program planned at Jefferson Lab to search for them.

## 5b.1 Theory and Motivation

The experiments at Jefferson Lab are particularly suited to probing $A'$ masses in the MeV-GeV range, which arise naturally in the models of Refs. [5-26, 5-27, 5-28, 5-29]. This $A'$ mass range is particularly intriguing. An $A'$ could explain the discrepancy between the measured and calculated value of the anomalous magnetic moment of the muon [5-30]. This long-standing discrepancy provides a particularly compelling experimental "target" shown with a green band in *Figure 5.6*. It could also explain some of the data of cosmic-ray, balloon-borne, or terrestrial dark matter experiments [5-31, 5-32]. Dark matter, which consists of some unknown particle(s) and makes up 80% of the matter density in our Universe, could be part of a hidden sector and couple to an $A'$. This would produce new dark matter interactions with ordinary matter and might explain some of the observed data "anomalies". For example, a rise in the observed number of cosmic-ray positrons [5-33] at energies of ~ 10 GeV and higher, above what is naively expected from astrophysical background sources suggests a contribution from dark matter annihilation into $A'$s, which then decay to electron-positron pairs. Intriguingly, the interactions of dark matter with an $A'$ can also provide the required annihilation rate.

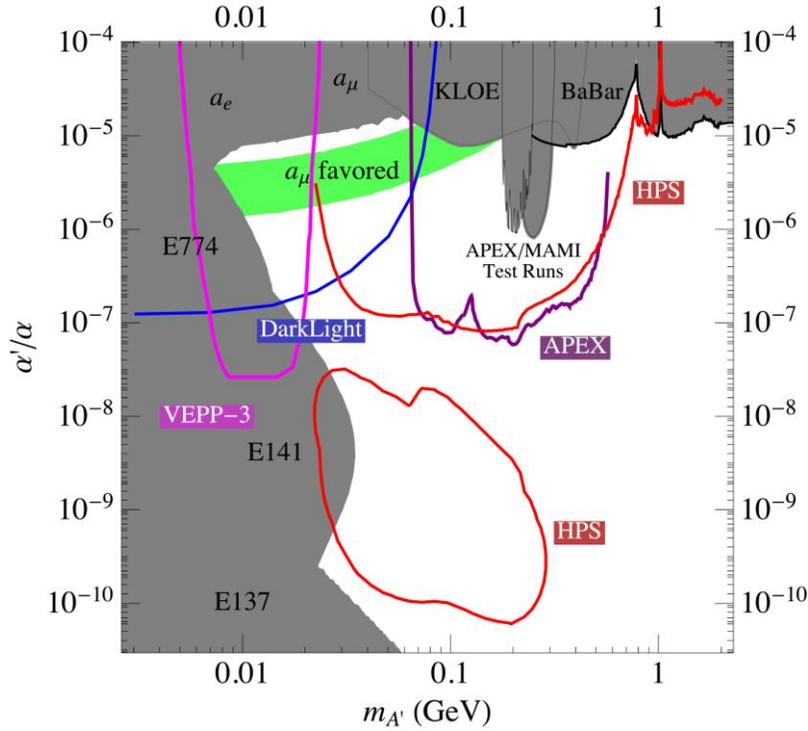

*Figure 5.6: Existing constraints and planned searches for heavy photons (A'). Shown are existing 90% confidence level limits from the beam dump experiments E137, E141, and E774 [5-34, 5-35, 5-36, 5-37], the muon anomalous magnetic moment $a_\mu$ [5-34, 5-59], KLOE [5-38], the test run results reported by APEX [5-39] and MAMI [5-40], an estimate using a BaBar result [5-34, 5-41, 5-42], and a constraint from supernova cooling [5-34] (see also [5-43]). In the green band, the A' can explain the observed discrepancy between the calculated and measured muon anomalous magnetic moment [5-30] at 90% confidence level. Projected sensitivities are shown for the full APEX run [5-43], HPS [5-45], Dark Light [5-46], and VEPP-3 [5-47]. MAMI has plans (not shown) to probe similar parameter regions as these experiments. Existing and future $e^+e^-$ colliders like BABAR, BELLE, KLOE, SuperB, BELLE-2, and KLOE-2 can also probe large parts of the parameter space for $\alpha'/\alpha \gtrsim 10^{-4} - 10^{-3}$ (not shown).*

Existing constraints [5-34] and the sensitivity of several planned experiments are shown in **Figure 5.6** in a plot of $\alpha'/\alpha$ (the ratio of the $A'$ coupling to that of QED) and the mass of the $A'$. We now discuss experimental searches for the $A'$, with particular focus on the electron fixed target experiments planned at Jefferson Lab.

## 5b.2 Experimental Searches at Jefferson Lab

An $A'$ in the MeV-GeV mass range can decay to electrically charged particles (e.g. $e^+e^-, \mu^+\mu^-, or\, \pi^+\pi^-$) or to light hidden-sector particles (if available), which can in turn decay to ordinary matter. Such an $A'$ can be efficiently produced in electron- or proton-fixed-target experiments [5-34, 5-39, 5-40, 5-44, 5-45, 5-46, 5-47, 5-48, 5-49] and at $e^+e^-$ and hadron colliders [5-17, 5-27, 5-38, 5-41, 5-50, 5-51, 5-52, 5-53, 5-54, 5-55, 5-56, 5-57, 5-58].

Electron fixed-target experiments are well suited to probe a large range in the $\alpha'/\alpha - m_{A'}$ parameter space [5-34]. In particular, the large luminosity ($\mathcal{O}(1\ ab^{-1}/$ day$))$ presently available at CEBAF and the Free Electron Laser (FEL) at Jefferson Lab, and their beam characteristics make Jefferson Lab particularly well suited for $A'$ searches.

In electron fixed-target experiments, the $A'$ is produced via bremsstrahlung from the incoming electron beam as it interacts with the target nuclei, see **Figure 5.7**. The Jefferson Lab experiments are sensitive to $A'$ decays to an $e^+e^-$ pair (or in some cases also a $\mu^+\mu^-$ pair). This decay can occur promptly or produce an $\mathcal{O}$(cm) displaced vertex if $\alpha'$ is sufficiently small.

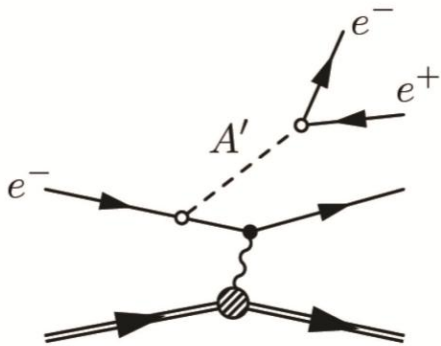

*Figure 5.7: Reaction mechanism for A' production via bremsstrahlung in the interaction of an electron with a nuclear target.*

Radiative and Bethe-Heitler trident production give rise to large backgrounds, and three experimental approaches have been proposed to detect the $e^+e^-$ pair from the $A'$ decay over these backgrounds: dual-arm spectrometers (like APEX) search for a small resonance in the $e^+e^-$ invariant mass spectrum on top of a large smooth background; forward vertexing spectrometers (like HPS) search for a small vertex from the finite $A'$ lifetime; and full final state reconstruction in experiments (like DarkLight) detect all outgoing particles to infer the presence of the $A'$. The complementary approaches map out different regions in the mass-coupling parameter space.

APEX [5-44] in Hall A is a proposed experiment that would use the CEBAF electron beam at various energies incident on a tungsten target. The $e^+e^-$ pairs produced from an $A'$ decay can be detected using the existing High Resolution Spectrometers and the septum magnet in Hall A. A successful test run in Hall A demonstrated the feasibility of this approach [5-39]. HPS [5-45] is a proposed experiment sited in Hall B, downstream of the CLAS12 detector. It would utilize a low current electron beam at 4-11 GeV on a tungsten target. Its unique feature is the capability to detect pairs from $A'$ decays producing a vertex downstream of the target and the ability to also detect muon pairs. The DarkLight experiment [5-46] would extend the search for an $A'$ to lower masses. It would utilize the high intensity electron beam at 140 MeV available at the FEL facility at Jefferson Lab, incident on a diffuse gas hydrogen target.

In these ways the Jefferson Lab program will delineate, or at least set world-leading limits, on the space for new physics beyond the Standard Model in electroweak interactions.

# 6   Appendix A:  Experimental Equipment

## Overview

**The experimental program using the beam from the upgraded 12 GeV CEBAF** will address a broad range of nuclear and Standard Model physics topics reviewed in sections 1-5. These require a variety of experimental approaches.  A suite of spectrometers and detectors to support the planned experimental program is being prepared as a part of the 12 GeV upgrade to CEBAF. In two of the existing halls new spectrometers are added, a large acceptance-device called CLAS12 in Hall B and a precision magnetic spectrometer called Super High Momentum Spectrometer, or SHMS, in Hall C. The new experimental Hall D will make use of a tagged bremsstrahlung photon beam and solenoidal detector to house the GlueX experiment. Upgrades to the beam line and associated beam polarimetry in Hall A gives flexibility to include novel one-of-a-kind large-installation experiments. The four experimental Halls together will produce a core set of capabilities, which also allow a natural upgrade path as the experimental program undertakes more demanding and exclusive experiments after the initial suite of approved efforts.

The foreseen equipment in the four halls is well matched to the demands of the measurements described in sections 2-5.  The initial program in Hall A makes use of both the existing High Resolution Spectrometers and the planned Super BigBite Spectrometer for a program of complementary high-luminosity nuclear structure studies and nucleon form factor studies, the latter allowing a transverse spatial quark flavor decomposition to distance scales deep inside the nucleon. Hall A will also be used for special setup experiments, such as the Møller experiment and the SoLID experiment. The upgraded Hall B spectrometer will be suited in particular for understanding nuclear structure via generalized parton distributions and for studying propagation of quarks in (cold) nuclear matter.  These measurements require detection of several final state particles and good missing-mass resolution or observation of exclusive final states, accomplished by the very large acceptance of the Hall B multi-particle spectrometer which spans laboratory polar angles from 5 to 140 degrees.  In Hall C the coincidence magnetic spectrometer setup supports high-luminosity experiments detecting reaction products with momenta up to the full beam energy, a virtue well matched for making precision determinations of the valence quarks in nucleons and nuclei. The new Hall D and its dedicated spectrometer are designed to search for QCD-predicted hybrid mesons using photoproduction via linearly polarized photons, which in turn enables a partial wave analysis of any new observed states to determine whether they have allowed or exotic quantum numbers.  The spectrometer can also perform studies of meson and baryon spectroscopy, which bear on the physical origins of quark confinement.

Details of the different spectrometers are given in the following:

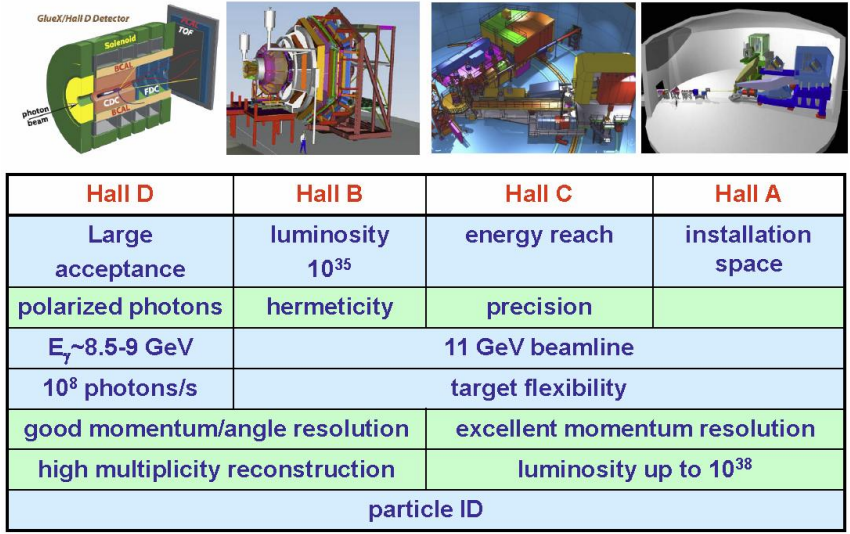

| Hall D | Hall B | Hall C | Hall A |
|---|---|---|---|
| Large acceptance | luminosity $10^{35}$ | energy reach | installation space |
| polarized photons | hermeticity | precision | |
| $E_\gamma$~8.5-9 GeV | 11 GeV beamline | | |
| $10^8$ photons/s | target flexibility | | |
| good momentum/angle resolution | excellent momentum resolution | | |
| high multiplicity reconstruction | luminosity up to $10^{38}$ | | |
| particle ID | | | |

*Figure 6.1 Comparison of major features of the four experimental Halls*

The planned experiments vary in required luminosity by more than three orders of magnitude and have momentum resolution requirements ranging from a few percent to $10^{-4}$. Some experiments require space and precise magnetic fields to employ polarized targets. Parity violation experiments require polarized electrons, which are routinely produced already at CEBAF, and many of the electroproduction experiments planned, e.g. DVCS experiments also require polarized electrons. All will require excellent identification of electrons, and tagging and identification of scattered target nucleons, and/or produced hadrons such as pions or kaons. The range of requirements necessarily leads to major differences among the Halls and some specialization of capabilities. An overview is given in ***Figure 6.1***, which compares general features of the high luminosity Halls A and C with those with the large acceptance detectors in Halls B and D.

| Topic | Hall A | Hall B | Hall C | Hall D | Total |
|---|---|---|---|---|---|
| The Hadron spectra as probes of QCD (*GlueX* and heavy baryon and meson spectroscopy) | | 1 | | 1 | **2** |
| The transverse structure of the hadrons (Elastic and transition Form Factors) | 4 | 3 | 2 | | **9** |
| The longitudinal structure of the hadrons (Unpolarized and polarized parton distribution functions) | 2 | 2 | 5 | | **9** |
| The 3D structure of the hadrons (Generalized Parton Distributions and Transverse Momentum Distributions) | 5 | 10 | 3 | | **18** |
| Hadrons and cold nuclear matter (Medium modification of the nucleons, quark hadronization, N-N correlations, hypernuclear spectroscopy, few-body experiments) | 3 | 2 | 6 | | **11** |
| Low-energy tests of the Standard Model and Fundamental Symmetries | 2 | | | 1 | **3** |
| **TOTAL** | **16** | **18** | **16** | **2** | **52** |

*Table 6.1 Categories of Approved Experiments, grouped by Hall*

The construction schedule for the 12 GeV project includes planned dates for first beam and checkout in each Hall. This will be inter-mixed with engineering runs together with machine development periods to achieve physics quality beam. After the formal end of the 12-GeV construction project, foreseen in 2015, there is planned a period of continuous machine development and first physics running. The 12 GeV approved physics experiments to date are summarized in Table 6.1, which gives the number of experiments in six general categories that have already been approved for each of the four Halls.

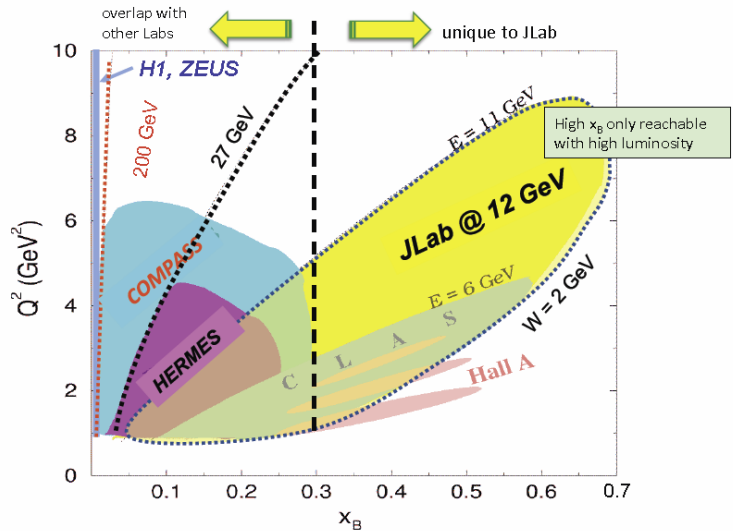

*Figure 6.2 Plot of kinematic coverage for the 12 GeV upgrade as a function of $Q^2$ vs. $x_B$, comparing coverage of other major facilities*

The kinematic coverage provided by the 12 GeV upgrade is shown in **Figure 6.2**. In contrast to efforts at HERA (H1, ZEUS) and CERN (COMPASS), which focus on the sea quark region below $x_B$ of 0.1-0.01, the 12 GeV CEBAF is particularly suited to examine the valence quark region of $x_B > 0.25$. The possible beam currents reach 100 microamps, which could be used in Halls A and C. It is seen in **Figure 6.3** that the 12 GeV Upgrade provides substantially enhanced access to the deep-inelastic scattering (DIS) regime with enough luminosity to reach the high-$Q^2$, high-$x_B$ region. The plot shown is for a luminosity typical of Hall B experiments; good count rates out to the limits are seen.

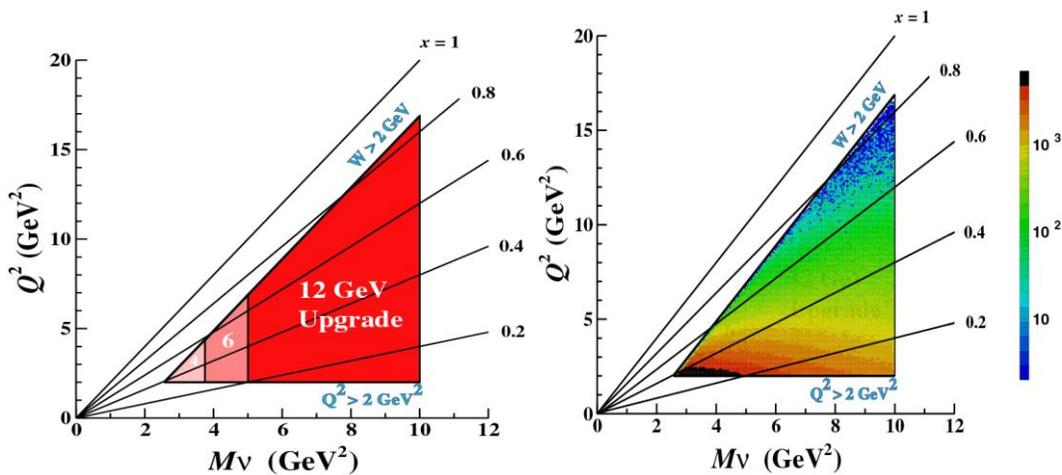

*Figure 6.3 Comparison of kinematic coverage of 12 GeV to 6 and 4 GeV CEBAF (left) and counts/hour at $L=10^{35}/cm^2$*

## Hall A

The existing High Resolution Spectrometers in Hall A are to be kept as part of the 12 GeV suite of equipment, and augmented with dedicated one-of-a-kind apparatus. An overview of the basic Hall layout, showing the vast 174-ft diameter space for specific large-instalation experiments, is shown in *Figure 6.4*. The existing HRS spectrometers access a momentum range of 0.3 to 4 GeV/c. One detector package is optimized for electrons, the other for hadrons. The Compton and Møller polarimeters are being upgraded to operate up to 11 GeV, the maximum energy planned for Halls A, B and C after the 12 GeV Upgrade. The magnets of the ARC energy measurement system will be refurbished and remapped at the larger excitations to provide beam energy measurement to $\Delta E/E = \pm 5 \times 10^{-4}$.

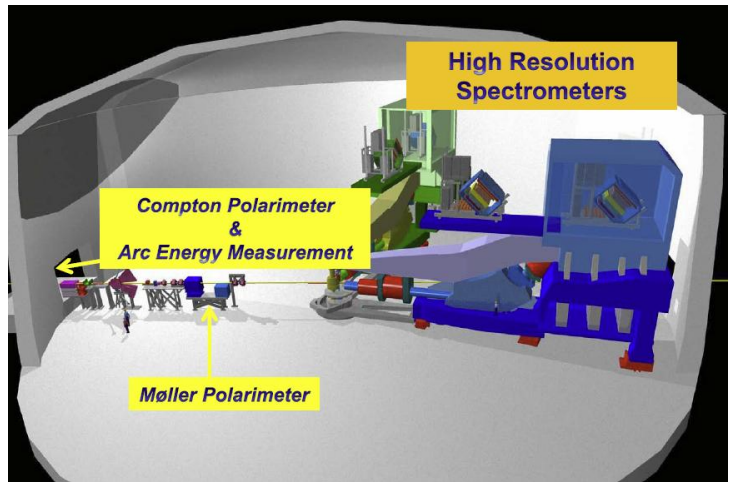

*Figure 6.4 Layout of Hall A showing the existing pair of High Resolution Spectrometers and Polarimeters being Upgraded*

Hall A extends its physics reach with a combination of general and specialized detector systems. A variety of devices, many of which have already been used, will continue to be available for experiments. These include septum magnets, the BigBite spectrometer, the BigCal calorimeter, the DVCS detector, the $G_E^n$ neutron detector, and potentially the existing HKS/HES spectrometers for hypernuclear spectroscopy studies. The capability of the Hall for one-of-a-kind experiments studying, for example, nucleon form factors, fundamental symmetry tests, and semi-inclusive physics is maintained; see the discussion below under New Initiatives.

## Hall B

The new CEBAF Large Acceptance Spectrometer 12 GeV (CLAS12) consists of two large detectors operating in concert, a forward one based on a new superconducting torus magnet and a central one with a new superconducting solenoid magnet. Broad coverage is needed to track forward-scattered particles and detect recoils and multiple produced particles over a large range of angles as encountered in SIDIS and DVCS running, and in studying quark propagation through cold nuclear matter and associated hadron production. Parts of the existing CLAS such as the electromagnetic (EM) calorimeter, part of a forward time of flight (TOF) wall, and the existing forward Cerenkov detector will be refurbished and re-used. A diagram of CLAS12 is shown in *Figure 6.5*.

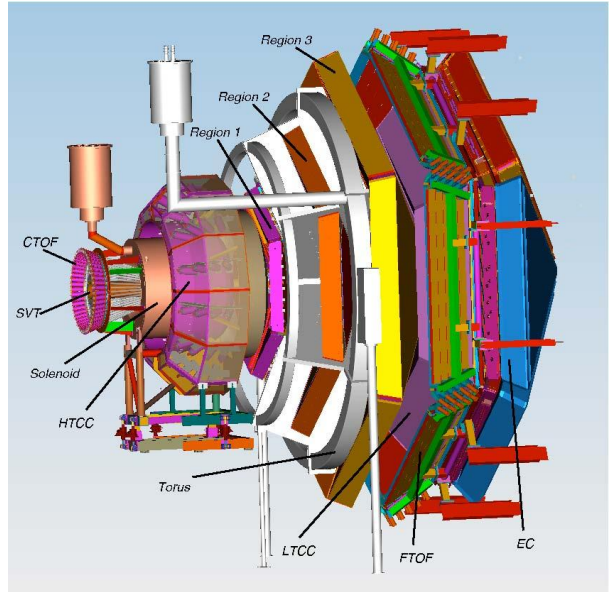

*Figure 6.5 Schematic of CLAS12*

CLAS12 is matched to the kinematics of the new 12-GeV-energy beams and can operate at 10 times the luminosity of the earlier CLAS. The torus will have 6-fold symmetry and cover polar angles from 5-40 degrees, with azimuthal coverage ranging from 50-90% of $2\pi$ depending on polar angle. The integrated magnetic field strength ranges from 2.3 Tm at 5 degrees to 0.53 Tm at 40 degrees. Three planes of drift chambers, each with two superlayers, will be integrated with each of the six sectors of the torus. Downstream of the drift chambers will be six identical sectors matched to the open acceptance of the torus magnet, each sector including a refurbished Cerenkov counter, followed by a new TOF wall and then a new pre-shower calorimeter placed just before the existing EM calorimeter.

The central detector area consists of a 5T ~1 meter long self-shielded solenoid with a 78 cm diameter bore, covering polar angles from 40 to 135 degrees. The central field is uniform to $10^{-4}$ in order to operate with a polarized target. The target is surrounded by a four-layer silicon-strip vertex tracker (SVT) arranged in a barrel configuration. Outside the SVT is a 48-element central TOF array, read out from both ends by photomultiplier tubes mounted on long light guides. Radial space and some downstream axial space in the solenoid bore is reserved for foreseen upgrades involving neutron detection and in particular additional tracking to identify detached vertices from hyperon decays. The space between the solenoid and the torus is occupied by a new Cerenkov detector. Particles passing to the forward spectrometer must traverse the thin mirror of this high-threshold Cerenkov counter (HTCC), with the entire HTCC contained in a thin-walled vessel operating at atmospheric pressure. A heavy tungsten Møller shield, which is supported from the torus, in combination with the 5T solenoid field prevents Møller electrons from overwhelming the spectrometer.

The entire CLAS12 is readout using TDCs, pipelined flash ADCs and a pipelined trigger technique similar to that noted below for Hall D.

# Hall C

The existing High Momentum Spectrometer is retained in Hall C and is supplemented by a new Super High Momentum Spectrometer (SHMS). This pair of heavily shielded magnetic spectrometers allows for high-precision measurements of neutrino-like cross sections to map valence quarks in nucleons and nuclei. The optical configuration of both magnetic spectrometers is similar, being QQQD for the HMS and being supplemented by a horizontal bender to reach angles as small as 5.5 degrees for SHMS, resulting in DQQQD.  The main Dipole bends vertically and transports accepted particles to a detector package in a heavily-shielded area on the spectrometer platform.  The shield includes inner linings of lead and boron-carbide to suppress neutrons and gamma rays, all encased in heavy concrete walls.  The boron carbide is included as a 'sand' in a pourable concrete form developed and patented at Jefferson Lab.  All five magnets are superconducting, bath-cooled by liquid helium, and with custom designs.  An overall layout for the SHMS is shown in *Figure 6.6*.  The foreseen detector package for SHMS mimics that of HMS.  Upon exit from the main dipole magnet particles traverse a noble-gas Cerenkov counter filled with Ar or Ne, then pass through a set of *x-y* drift chambers, then transit the first of two *x-y* scintillator hodoscopes used for rough trajectory determination plus triggering.  They next pass through a heavy gas Cerenkov counter, filled with $C_4F_8O$, then transit a second pair of *x-y* trigger hodoscopes, and finally end in an electromagnetic calorimeter including a pre-shower and shower section, both of lead glass blocks.  The downstream 'Y' hodoscope plane is made of cast quartz.  In between the downstream hodoscope pair and the pre-shower detector a space is reserved for specific detectors, most notably a pair of aerogel-based Cerenkov counters for kaon identification. The entire detector package sits on an optical bench to align it with the magnetic optical axis of the dipole.  The hodoscopes, Cerenkov counters, and calorimeters are read out by PMTs, while the drift chambers are read out by a standard electronics chain including amplifier, discriminator and TDC.

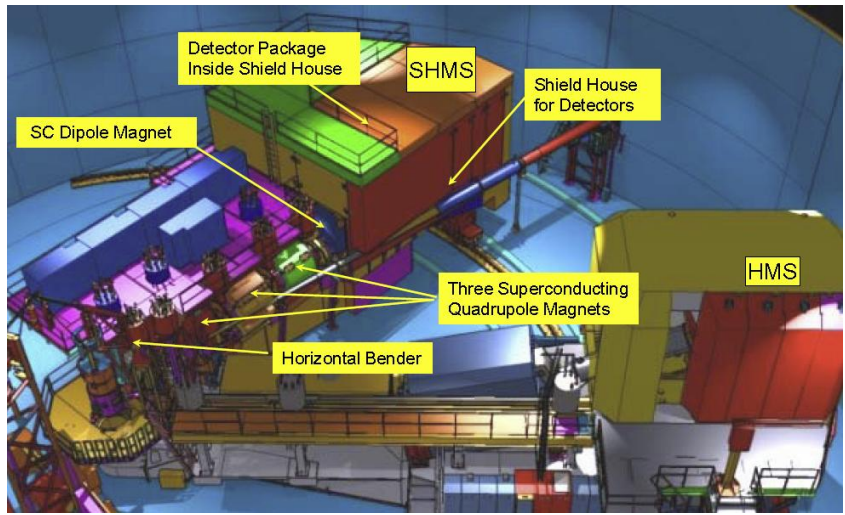

*Figure 6.6 Hall C layout showing SHMS (top) and existing HMS.*

The readout uses the pipelined ADCs developed for Hall D. Space is reserved in the detector 'stack' for future detectors, notably an aerogel-based Cerenkov counter. It is possible to remove some of the detectors and insert a proton polarimeter, for instance, as experiment needs dictate.

## Hall D

Hall D searches for new exotic states and strives to measure mass and $J^{PC}$ for them. This requires a nearly-hermetic detector to be able to reconstruct fully final states of mass up to 2.5 GeV/c$^2$ or higher. It also requires sufficient information to perform partial wave analyses, with the chosen method being to excite the states by means of linearly polarized photons. Some example final states are $\eta_1 \to a^+_1\pi^- \to (\rho^0\pi^+)(\pi^-) \to \pi^+\pi^-\pi^+\pi^-$, where all final state particles are charged, $h_0 \to b^0_1\pi^0 \to (\omega\pi^0)\gamma\gamma \to \pi^+\pi^-\gamma\gamma\gamma\gamma$, where there are multiple final state photons, and $h'_2 \to K^+_1 K^- \to \rho^0 K^+ K^- \to \pi^+\pi^- K^+ K^-$, requiring identifying strange particles.

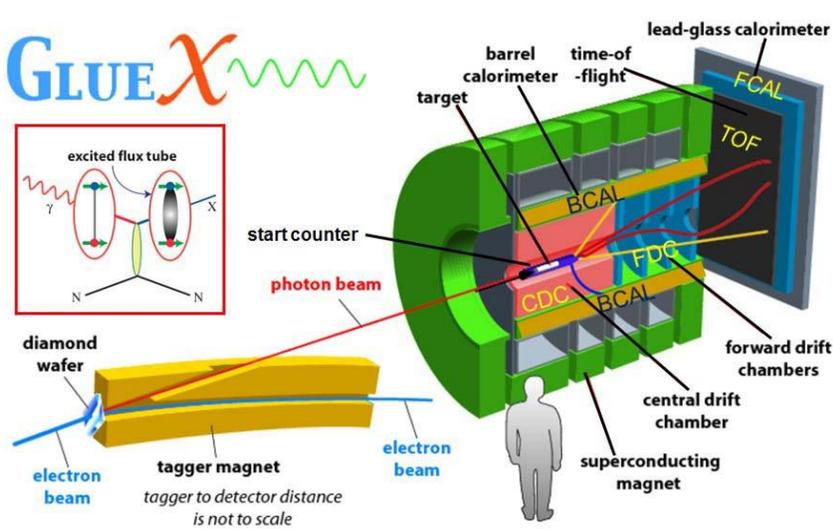

*Figure 6.7 Schematic of Hall D GlueX detector*

The detector for Hall D, known as the *GlueX* detector, is shown in *Figure 6.7*. The linearly polarized photons, some $10^7 - 10^8$/s, are produced by coherent bremsstrahlung in a 20 micron thick diamond crystal located 75 meters upstream of a 3mm diameter collimator. Scattered electrons are bent by a tagger magnet located just after the diamond and equipped with both a tagging hodoscope to cover a broad range of energies as well as a tagging microscope placed near the electron energy corresponding to the coherent bremsstrahlung peak near 9 GeV to determine the photon energy to 0.l%. The detector is based on a large 2T superconducting solenoid magnet of 2 meter bore and 4 meter length, with a target placed near the upstream end. A cylindrical tracking chamber is placed around the target and works in tandem with planar tracking chambers placed downstream. Both sets of chambers are enclosed by an annular Barrel electromagnetic Calorimeter, comprised of thin lead sheets interleaved with axial scintillating fibers. The light produced by the scintillating fibers is read out by newly-developed Silicon Photomultipliers, which being purely solid-state devices can operate in the 2T magnetic field without affect on their gain. Downstream of the solenoid is a TOF wall backed by a lead-glass electromagnetic calorimeter consisting of 2800 towers.

All elements are read out by fast digitizers, both ADCs and TDCs, which operate in a pipelined fashion. Both timing and pulse height information are taken from the calorimeters and TOF. Custom 250 (125) MHz flash ADC designs are used for the calorimeters and TOF walls (tracking chamber readout). The TDCs are based on an existing pipelined ASIC. This permits use of a pipelined Level-1 trigger operating in hardware and making selections based on for example calorimetric energy and track count, and an event-based Level-3 trigger operating in a computing farm and working on fully assembled events. A custom set of six types of trigger modules support trigger decisions using this pipelined architecture.

## Other Existing Equipment and Upgrades to Planned Base Equipment

A suite of existing targets is available for the 12 GeV program. These include liquid hydrogen and deuterium cryogenic targets, polarized $^3$He targets (to serve for example as a quasi-free neutron target), a new novel solid HD target that can also be polarized, a new NSF-supported longitudinally polarized target for CLAS12 constructed by users, and various other polarized targets used as part of the 6 GeV program at CEBAF. There are large acceptance detectors for recoiling and produced particles such as the Big Bite spectrometer used in Hall A and various electromagnetic calorimeter walls and neutron detector walls.

New devices are in various stages of development by outside groups to supplement the above detector suites. For Hall B these include a neutron detector, forward tracker, added barrel tracker sections, and a forward tagger. CLAS12 can also be supplemented by a new radial TPC around the target which will be used to study structure of bound nucleons. For Hall C these include a neutral-pion spectrometer facility, a large backward-angle nucleon wall, and a neutron polarimeter. A desired detector for both Halls B and D is a RICH counter to improve kaon identification. The required designs are rather different, with the Hall B one following the torus sector geometry and that for Hall D needing to integrate with the forward tracker.

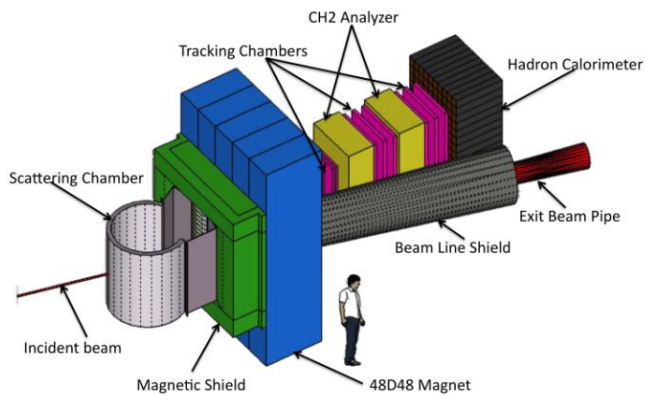

*Figure 6.8 Layout of SuperBigBite Spectrometer*

In Hall A the foreseen equipment includes a Super Big Bite spectrometer (SBS) to perform a series of elastic nucleon form factor measurements down to the smallest distance scales accessible at Jefferson Lab. Given the rapid fall-off of the form factors with $Q^2$ a high-luminosity device with large acceptance is needed. SBS is a tracking spectrometer built around a large acceptance dipole modified to include a slot in the yoke for beam transport to enable large aperture operation close to the beam axis. A set of some 64 gas electron multiplier (GEM) tracking chambers are needed and have been prototyped for SBS. These permit very high rate operation with good two-dimensional readout. A layout is shown in *Figure 6.8.*

## Future Initiatives

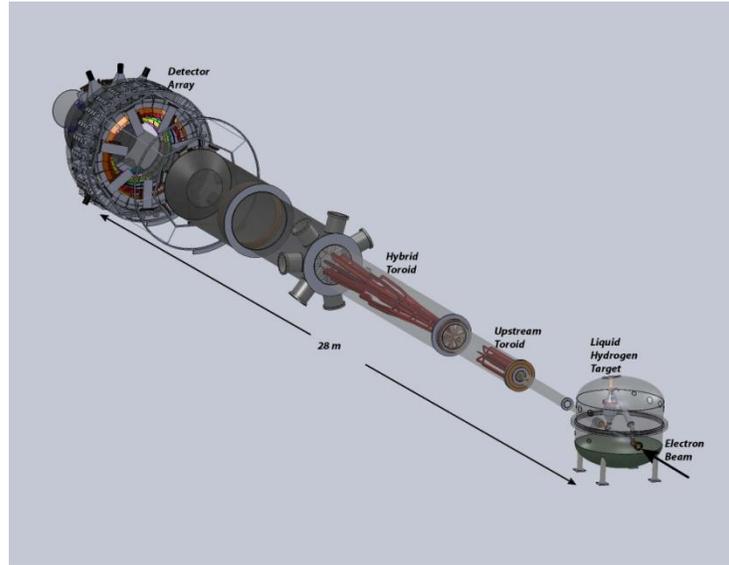

*Figure 6.9 The Layout of the MOLLER Experiment*

## MOLLER

The MOLLER experiment, whose physics goals were described in section 5.A.2, plans to measure a parity-violating asymmetry in fixed target electron scattering to unprecedented precision. The MOLLER apparatus is planned to be mounted in Hall A and has several novel features; see *Figure 6.9*. An 11 GeV electron beam would be incident on a 1.5 meter long 5 kW liquid hydrogen target. The foreseen apparatus will consist of two toroidal magnets, a design to preclude the need for dipoles immediately downstream of the target to remove background.

MOLLER will benefit from the steady improvement in the techniques employed to measure parity-violating asymmetries to sub-ppb systematic precision and to achieve normalization control at the sub-% level. Two redundant continuous monitors of electron beam polarization would be employed to achieve normalization control at 0.4%. Auxiliary detectors would track individual particles at low rates to measure the absolute value of $Q^2$ to 0.5%. Very forward angle detectors downstream of the main detectors would monitor potentially fatal luminosity fluctuations due to jitter in electron beam properties and target density. Several methods to reverse the sign of the asymmetry will also be employed periodically.

## SOLID

The SOLenoidal Large Intensity Device (SoLID) is a versatile apparatus based on a large solenoidal magnet that is proposed for two series of experiments. One arrangement involves measuring parity violating deep inelastic scattering (PVDIS) and would focus on lepton-quark neutral current interactions. The experiment requires 50 µA of 85% polarized electrons together with a large acceptance, which is achieved by using a large bore solenoid of 2-3 meter diameter and 1.5T field. Re-use of an existing large superconducting solenoidal magnet is being pursued. The detection system requires excellent electron identification which is achieved by use of Gas Electron Multiplier (GEM) tracking chambers coupled with gas Cerenkov counters and an electromagnetic calorimeter. A series of baffles would be deployed to block soft electrons' trajectories.

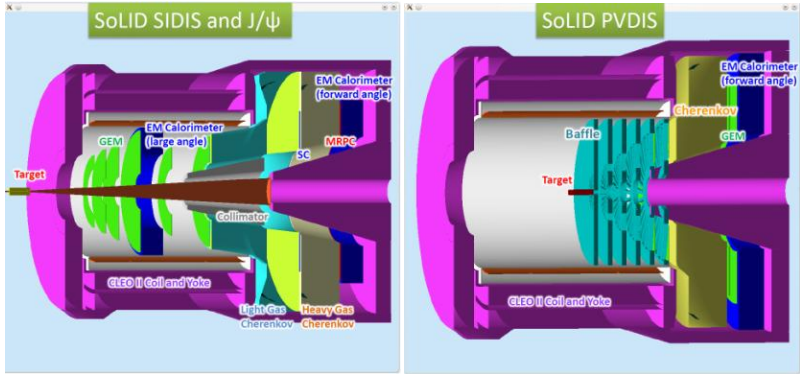

*Figure 6.10 Layout of SOLID Experiment in SIDIS and PVDIS*

A modified setup after removing baffles enables studies of semi-inclusive deep inelastic scattering (SIDIS) with optimal (nearly $2\pi$) azimuthal-angle coverage. Electron-hadron separation would be accomplished with a light gas Cerenkov and an electromagnetic calorimeter. Pion identification would be achieved by a heavy gas Cerenkov and a layer of multi-gap resistant plate chambers to determine time of flight.

The SIDIS setup will be also used to measure cross section of the electroproduction of J/$\psi$ mesons off a proton near threshold, which is sensitive to the gluonic interaction between the J/$\psi$ and the nucleon. Such a measurement would open a new window to study QCD in the non-perturbative regime (e.g. the conformal anomaly) using charmonium.

# Acknowledgement

Comments and suggestions from many scientists in the Jefferson Lab user community were extremely valuable in the course of the editing of this white paper, and are gratefully acknowledged. This work was supported by DOE contract DE-AC05-06OR23177, under which Jefferson Science Associates, LLC, operates the Thomas Jefferson National Accelerator Facility.

# Appendix B
## Approved 12 GeV Era Experiments to Date (July 2012)

| | |
|---|---|
| E12-06-101 | Measurement of the Charged Pion Form Factor to High $Q^2$ |
| E12-06-102 | Mapping the Spectrum of Light Quark Mesons and Gluonic Excitations with Linearly Polarized Photons |
| E12-06-104 | Measurement of the Ratio R = $\sigma_L/\sigma_T$ in Semi-Inclusive DIS |
| E12-06-105 | Inclusive Scattering from Nuclei at *x* > 1 in the quasi-elastic and deep-inelastic regimes |
| E12-06-106 | Study of Color Transparency in Exclusive Vector Meson Electroproduction off Nuclei |
| E12-06-107 | The Search for Color Transparency at 12 GeV |
| E12-06-108 | Hard Exclusive Electroproduction of $\pi^0$ and η with CLAS12 |
| E12-06-109 | The Longitudinal Spin Structure of the Nucleon |
| E12-06-110 | Measurement of Neutron Spin Asymmetry $A_1^n$ in the Valence Quark Region Using an 11 GeV Beam and a Polarized $^3$He Target in Hall C |
| E12-06-112 | Probing the Proton's Quark Dynamics in Semi-Inclusive Pion Production at 11 GeV |
| E12-06-113 | The Structure of the Free Neutron at Large x-Bjorken |
| E12-06-114 | Measurement of Electron-Helicity Dependent Cross Sections of Deeply Virtual Compton Scattering with CEBAF at 12 GeV |
| E12-06-117 | Quark Propagation and Hadron Formation |
| E12-06-119 | Deeply Virtual Compton Scattering with CLAS at 11 GeV |
| E12-06-121 | A Path to "Color Polarizabilities" in the Neutron: A Precision Measurement of the Neutron $g_2$ and $d_2$ at High $Q^2$ in Hall C |
| E12-06-122 | Measurement of the Neutron Asymmetry $A_1^n$ in the Valence Quark Region using 8.8 and 6.6 GeV Beam Energies and BigBite spectrometer in Hall A |
| E12-07-104 | Measurement of the Neutron Magnetic Form Factor at High $Q^2$ Using the Ratio Method on Deuterium |
| E12-07-105 | Scaling Study of the L-T Separated Pion Electro-production Cross-Section at 11 GeV |
| E12-07-107 | Studies of Spin-Orbit Correlations with Longitudinally Polarized Target |
| E12-07-108 | Precision Measurement of the Proton Elastic Cross Section at High $Q^2$ |
| E12-07-109 | Large Acceptance Proton Form Factor Ratio Measurements at 13 and 15 (GeV/c)$^2$ Using Recoil Polarization Method |
| E12-09-002 | Charge Symmetry Violating Quark Distributions via Precise Measurement of $\pi^+/\pi^-$ Ratios in Semi inclusive Deep Inelastic Scattering. |
| E12-09-003 | Nucleon Resonance Studies with CLAS |
| E12-09-005 | An Ultra-Precise Measurement of the Weak Mixing Angle using Moeller Scattering |

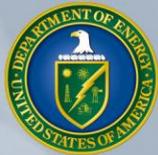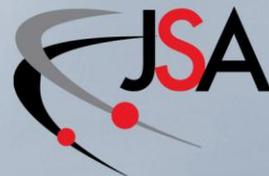